\documentclass[12pt,draftclsnofoot,onecolumn,twoside]{IEEEtran}
\usepackage[mathscr]{eucal}
\usepackage{ifpdf}
\usepackage{cite}
\usepackage{graphicx}
\usepackage[cmex10]{amsmath}
\usepackage{amssymb}
\usepackage{algorithmic}
\usepackage{units}
\usepackage{setspace}
\usepackage{algorithm}
\usepackage{array}
\usepackage[tight,footnotesize]{subfigure}
\usepackage{amsthm}
\usepackage{multirow}
\usepackage{enumerate}
\usepackage{color}

\newcounter{assume}
\def\theassume{(A\arabic{assume})}
\newtheorem{theorem}{Theorem}
\newtheorem{lemma}{Lemma}
\newtheorem{proposition}{Proposition}
\newtheorem{corollary}{Corollary}

\newtheorem{remark}{Remark}

\newtheorem{definition}{Definition}

\usepackage{cancel}
\usepackage{soul}
 \newcommand{\resl}[1]{}
\newcommand{\rema}[1]{}
 \newcommand{\tosl}[1]{}
 \newcommand{\cosl}[1]{}
 \newcommand{\insl}[1]{#1}

\begin{document}
\title{Joint Channel Selection and Power Control in Infrastructureless
  Wireless Networks: A Multi-Player Multi-Armed Bandit Framework}
%
\author{
\IEEEauthorblockN{Setareh Maghsudi and S\l awomir Sta\'{n}czak, \textit{Senior Member, IEEE}\\}

\thanks{Parts of the material in this paper were presented at the IEEE Wireless Communications 
and Networking Conference, Shanghai, April, 2013. The work was supported by the German Research 
Foundation (DFG) under grant STA 864/3-3. The authors are with the Fachgebiet f\"ur Informationstheorie 
und theoretische Informationstechnik, Technische Universit\"at Berlin. The second author is also with 
the Fraunhofer Institute for Telecommunications Heinrich Hertz Institute, Berlin, Germany (e-mail: 
setareh.maghsudi@tu-berlin.de, slawomir.stanczak@hhi.fraunhofer.de).}%
}
\maketitle
%
%
\begin{abstract}
This paper deals with the problem of efficient resource allocation in dynamic 
infrastructureless wireless networks. Assuming a reactive interference-limited 
scenario, each transmitter is allowed to select one frequency channel (from a 
common pool) together with a power level at each transmission trial; hence, for 
all transmitters, not only the fading gain, but also the number of interfering 
transmissions and their transmit powers are varying over time. Due to the absence 
of a central controller and time-varying network characteristics, it is highly 
inefficient for transmitters to acquire global channel and network knowledge. 
Therefore a reasonable assumption is that transmitters have no knowledge of 
fading gains, interference, and network topology. Each transmitting node 
selfishly aims at maximizing its average reward (or minimizing its average 
cost), which is a function of the action of that specific transmitter as well 
as those of all other transmitters. This scenario is modeled as a multi-player 
multi-armed adversarial bandit game, in which multiple players receive an a 
priori unknown reward with an arbitrarily time-varying distribution by sequentially 
pulling an arm, selected from a known and finite set of arms. Since players 
do not know the arm with the highest average reward in advance, they attempt 
to minimize their so-called regret, determined by the set of players' actions, 
while attempting to achieve equilibrium in some sense. To this end, we design 
in this paper two joint power level and channel selection strategies. We prove 
that the gap between the average reward achieved by our approaches and that 
based on the best fixed strategy converges to zero asymptotically. Moreover, 
the empirical joint frequencies of the game converge to the set of correlated 
equilibria. We also characterize this set for two special cases of our designed 
game. We further discuss experimental regret-testing procedure as another 
potential solution, which converges to Nash equilibrium. Finally all approaches 
are compared through extensive numerical analysis.

%
\end{abstract}
\begin{keywords}
Adversarial bandits, channel selection, equilibrium, infrastructureless wireless network, power control.
\end{keywords}

\section{Introduction}\label{sec:Introduction}
%
\subsection{Bandit Theory and Wireless Communication}\label{subsec:BanditInt}
Multi-armed bandit (MAB) is a class of sequential optimization
problems, to the best of our knowledge originally introduced in 
\cite{Robbins52}. In the most traditional form of MAB, given a 
set of arms (actions), a player pulls an arm at each trial of 
the game to receive a reward. The rewards of arms are not known 
to the player in advance; however, upon pulling an arm, its 
instantaneous reward is revealed. In such unknown setting,
after playing an arm, the player may lose some reward (or incur
additional cost) due to not playing another arm instead of the 
currently played arm. This can be quantified by the difference 
between the reward that would have been achieved had the player 
selected another arm, and the reward of the played arm. This 
quantity is called \textit{regret}. The player decides which 
arm to pull in a sequence of trials so that its accumulated
regret over the game horizon is minimized. Such problems obviously
render the intrinsic trade-off between exploration (learning) 
and exploitation (control), i.e. playing the arm which has exhibited the
best performance in the past and playing other arms to guarantee the
optimal payoff in future. An important class of bandit games is
adversarial bandits, where the series of rewards generated by an arm
cannot be attributed to any specific distribution function.

In recent years, bandit theory has been used in communication theory. 
For instance, \cite{Liu10} and \cite{Liu12} utilize the classical 
bandit game to model spectrum sharing in cognitive radio networks. 
In \cite{Felice11}, the authors propose a cooperative spectrum sensing 
scheme based on bandit theory. Further, References \cite{Maghsudi13}, 
\cite{Krishnamurthy07}, and \cite{Nino11} use bandit theory to model 
relay selection, sensor scheduling and object tracking, respectively. 
Channel monitoring using bandit model is investigated in \cite{Arora11} 
and \cite{Zheng13}. Bandit models have been also used to solve the 
distributed resource allocation problem, as discussed in the following.

\subsection{Distributed Resource Allocation in Infrastructureless Wireless Networks}\label{subsec:RelatedWorks}
In recent years, game theory and reinforcement learning have been widely used to solve 
the distributed resource allocation problem. The vast majority of game-theoretic approaches 
are based on either cooperation (e.g. coalition formation), mechanism design (e.g. auction 
theory), or exchange economy (e.g. supply-demand markets). Although these approaches can 
be implemented in a distributed manner, such an implementation in a real network environment 
requires that each player at least knows its own utility function a priori. On the other 
hand, these approaches are in general inefficient as players have to exchange information 
for coordination, which increases signaling and feedback overhead. For example, most models 
from cooperative game theory require coordination and/or communication among players to 
construct coalitions \cite{Chen11}, \cite{Saad11}. In wireless resource allocation using 
auction games, bids must be submitted to some central controller that performs necessary 
computations and makes decisions \cite{Mukherjee10}, \cite{Sun06}. Finally, in supply-demand 
market models, prices and demands are exchanged among buyers and sellers \cite{Ileri05}, 
\cite{Maghsudi11}.

When the utility functions are not known in advance, the resource allocation problem is often 
solved by using learning approaches, including bandit models. A large body of literature, such 
as \cite{Guo04}, \cite{Fang13} and \cite{Song07}, analyze single-agent stochastic learning 
problems. Another example is \cite{Gai12}. In this work, network optimization is modeled as a 
stochastic bandit game, where at each trial multiple arms are selected by a single player and 
the reward is some linear combination of the rewards of selected arms. An application of this 
formulation might be a downlink user selection, performed by the base station. In single-agent 
settings, the agent learns from its previous experiences, and no information flow is required. 
However, this type of learning cannot generally be used in wireless networks, where multiple 
players act selfishly by responding to each other and their utilities are influenced by the actions of 
other players. Moreover, similar to games with complete information, it is desired that players 
achieve equilibrium in some sense. As for multi-agent settings, most studies assume that players 
are able to observe the actions of each other. This assumption, despite being realistic for 
some spectrum sharing problems, is not always applicable to general resource allocation problems, 
especially in power control games, where it is difficult to identify the transmit power level of 
players. In addition, the assumption that each player announces its actions (e.g. its transmit 
power) is not intensive compatible. As a result, a great majority of previous works focus on spectrum 
sharing and/or sensing, as well as channel monitoring. On the other hand, most of previous 
studies assume that the rewards achieved by each action can be attributed to a single density 
distribution. However this assumption is highly restrictive especially for dynamic networks.  

In \cite{KalathilD12}, multi-agent bandit problem is investigated. This study assumes that in 
case of interference, no reward is paid to interfering users, thereby eliminating interference, which degrades 
the overall performance depending on utility functions. In addition, communication among players 
is necessary. Finally, no equilibrium analysis is performed. Another example is Reference \cite{XuA12}, 
where opportunistic spectrum access is formulated as a multi-agent learning game. In this work, 
upon availability, each channel pays the same reward to all users so that this scenario is strictly 
restrictive as it neglects different channel qualities. Moreover, if a channel is selected by 
multiple users, orthogonal spectrum access scheme is used, which is known to be sub-optimal in 
general. References \cite{XuO13} and \cite{XuS13} consider graphical games for an interference 
minimization problem with partially overlapping channels, where the interference is present only 
between neighboring users. These works establish the convergence of proposed learning approaches 
for the special case of exact potential games; Nonetheless the analysis does not hold for more 
general games. The authors of \cite{Anandkumar10} model the cooperative rate maximization in cognitive 
radio networks as bandit game, and propose two approaches, depending on the availability of 
information. The stability of the solution is however not investigated. Reference \cite{Liang12} 
proposes two approaches that achieve Nash equilibrium in a multi-player cognitive environment. 
System verification, however, is only based on numerical approaches. References \cite{LiuS10}, 
\cite{LiuN10} and \cite{LiuM13} propose various selection schemes to achieve logarithmic regret; 
however, no equilibrium analysis is performed. All of the works named above assume that the generated 
rewards of any given action are independent and identically distributed.

\subsection{Our Contribution}\label{subsec:Contribution}
As discussed in Section \ref{subsec:RelatedWorks}, the resource
allocation problem using machine learning theory has been subject 
to extensive research in recent years. In short, our focus is 
on a resource allocation problem in an infrastructureless network. 
First, we model this problem as an adversarial multi-player multi-armed
bandit game. With the aim of an efficient management of network
resources and the co-channel interference mitigation, we follow an
approach suggested in \cite{BlumM07} to design two joint power control
and channel selection (PC-CS, hereafter) strategies, which are adapted
versions of \textit{exponential-based weighted average} \cite{Auer03}
and \textit{follow the leader} \cite{Kujala05} strategies. Both PC-CS
strategies not only result in small (that is, with sublinear growth
rate in time) regret for each individual player, but also guarantee
the convergence of empirical frequencies of play to the set of correlated 
equilibria. We further characterize this set for two special cases of 
our designed game. Moreover, we implement the \textit{experimental 
regret-testing procedure} \cite{Germano07}, which is shown to converge 
to the set of Nash equilibria of the game.

Our work extends the state-of-the-art in this area significantly since 
it differs from the existing studies in the following crucial aspects:

\begin{itemize}
\item We analyze the multi-agent bandit problem and take into account the 
selfishness of players.
\item We do not assume that the reward generating process of any given action 
is time-invariant. In fact, the reward functions are allowed to vary arbitrarily, 
which enables us to accommodate the dynamic nature of wireless channels and 
distributed networks.
\item We do not allow any communication among players, thereby minimizing 
the overhead. Moreover, players do \textit{not} observe the actions of 
each other, so that the developed model can be applied to a large body 
of resource allocation problems. An example is a power control problem 
with unknown power levels used by other players. We study a two-dimensional 
problem, namely joint channel and power level selection problem, by modeling 
it as a multi-player multi-armed bandit game. In our model, channel qualities 
are taken into account so that channels pay different rewards to different 
users. In addition, we impose no limitations on interference pattern.
\item Our convergence analysis is valid for a wide range of games. This 
is in contrast to many previous works where the game should be 
necessarily potential for the convergence analysis to hold.
\item We characterize the set of correlated equilibria for two special cases of our 
formulated game model.
\end{itemize}
\subsection{Paper Structure}\label{subsec:Organization}
Section \ref{sec:Preliminaries} briefly reviews some concepts and
results of bandit theory. In Section \ref{sec:System} the resource
allocation game is formulated. Section \ref{sec:BEBWA} presents a
PC-CS strategy based on \textit{exponential-based weighted average} 
rule \cite{Auer03}. In Section \ref{sec:BFPL}, another PC-CS strategy, 
derived from \textit{follow the leader} rule \cite{Kujala05} is discussed. 
Section \ref{sec:BERT} is devoted to \textit{experimental regret-testing} 
procedure \cite{Germano07}. Numerical analysis are presented in Section
\ref{subsec:Numerical}. Section \ref{sec:conclusion} concludes the paper.
\section{Multi-Player Multi-Armed Bandit Games}\label{sec:Preliminaries}
\subsection{Notions of Regret}\label{subsec:Regret}
Multi-player multi-armed bandit problem (MP-MAB, hereafter) is a class 
of sequential decision making problems with limited information. In 
this game, each player $k \in \left \{1,...,K \right \}$ is assigned 
an action set including $N_{k}$ actions (arms), $1\leq N_{k}\leq N$. 
Every player selects an action at successive trials in order to 
receive an initially unknown reward, which is determined not only by 
its own actions, but also by those of other players. The action set, 
the played action and the reward achieved by each player are regarded 
as private information. The reward generating processes of arms are 
independent. Let $\mathbf{I}$ and $I^{(k)}$ be the joint action space 
and the action space of player $k$, respectively. Accordingly, 
$\mathbf{I}_{t}= (I_{t}^{(1)},...,I_{t}^{(k)},...,I_{t}^{(K)})$ denotes 
the joint action profile of players at time $t$, with $I_{t}^{\left (k 
\right)}$ being the action of player $k$. Moreover, let $g_{t}^{(k)}(\mathbf{I}_{t}) 
\in [0,1]$ be the reward achieved by some player $k$ at time $t$.\footnote{Note 
that all results can be also expressed in terms of loss ($d$), provided 
that the loss is related to the gain by $d=1-g,\insl{g\in[0,1]}$.} The 
instantaneous regret of any player $k$ is defined as the difference 
between the reward of the optimal action,\footnote{Optimality is defined 
in the sense of the highest instantaneous reward.}~and that of the played 
action. Based on this definition, the cumulative regret of player $k$ is 
formally defined in the following.
\begin{definition}
\label{de:CumRegret} 
The cumulative regret of player $k$ up to time $n$ is defined as
\begin{equation}
\label{eq:CumRegret}
R_{n}^{(k)}= \max_{i=1,...,N_{k}} \sum_{t=1}^{n}g_{t}^{(k)}(i,\mathbf{I}_{t,k}^{-})-
\sum_{t=1}^{n}g_{t}^{(k)}(I_{t}^{(k)},\mathbf{I}_{t,k}^{-}),
\end{equation}
where $\mathbf{I}_{t,k}^{-}$ is defined to be the joint action
profile of all players except for $k$ at time $t$.
\end{definition}
Each player aims at minimizing its accumulated regret, which is an instance 
of the well-known exploitation-exploration dilemma: Find a desired balance 
between exploiting actions that have exhibited well performance in the past 
(control) on the one hand, and exploring actions which might lead to a better 
performance in the future (learning) on the other hand.

Now, suppose that players use mixed strategies. This means that, at each trial 
$t$, player $k$ selects a probability distribution $\mathbf{P}_{t}^{(k)}=
(p_{1,t}^{(k)},...,p_{i,t}^{(k)},...,p_{N_{k},t}^{(k)})$ over arms, and plays 
arm $i$ with probability $p_{i,t}^{(k)}$. In this case, we resort to expected 
regret, also called \textit{external regret} \cite{Bianchi06}, defined as follows.
\begin{definition}
\label{de:External}
The external cumulative regret of player $k$ is defined as
\begin{equation}
\label{eq:external}
\begin{aligned}
R_{\textup{Ext}}^{(k)}:=R_{\textup{Ext}}^{(k)}(n)=\max_{i=1,...,N_{k}}\sum_{t=1}^{n}
g_{t}^{(k)}(i,\mathbf{I}_{t,k}^{-})&-\sum_{t=1}^{n}\bar{g}_{t}^{(k)}\left (\mathbf{P}_{t}^{(k)},
\mathbf{I}_{t,k}^{-} \right )\\ 
=\max_{i=1,...,N_{k}} \sum_{t=1}^{n}\sum_{j=1}^{N}p_{j,t}^{(k)} & \left (g_{t}^{(k)}
(i,\mathbf{I}_{t,k}^{-})-g_{t}^{(k)}(j,\mathbf{I}_{t,k}^{-}) \right),
\end{aligned}
\end{equation}
where $\bar{g}_{t}^{(k)}(\cdot)$ denotes the \textit{expected} reward at round $t$ 
by using mixed strategy $\mathbf{P}_{t}^{(k)}$, defined as $\bar{g}_{t}^{(k)}(\cdot)=
\sum_{j=1}^{N}g_{t}^{(k)}(\cdot)p_{j,t}^{(k)}$. 
\end{definition}
By definition, external regret compares the expected reward of the current 
mixed strategy with that of the best fixed action in the hindsight, but 
fails to compare the rewards achieved by changing actions in a pair-wise 
manner. In order to compare actions in pairs, \textit{internal regret} \cite{Bianchi06} 
is introduced that is closely related to the concept of equilibrium in games.
\begin{definition}
\label{de:Internal}
The internal cumulative regret of player $k$ is defined as
\begin{equation}
\label{eq:internal}
\begin{aligned}
R_{\textup{Int}}^{(k)}:=R_{\textup{Int}}^{(k)}(n)=\max_{i,j=1,...,N_{k}} & R_{\left (i \to j \right),n}^{(k)}\\
=\max_{i,j=1,...,N_{k}} &\sum_{t=1}^{n}p_{i,t}^{(k)}\left (g_{t}^{(k)}\left (j,\mathbf{I}_{t,k}^{-}\right)- 
g_{t}^{(k)}\left (i,\mathbf{I}_{t,k}^{-} \right) \right).
\end{aligned}
\end{equation}
\end{definition}
Notice that on the right-hand side of (\ref{eq:internal}), $r_{(i \to j),t}^{(k)}=
p_{i,t}^{(k)}\left(g_{t}^{(k)}(j,\cdot)-g_{t}^{(k)}(i,\cdot)\right)$
denotes the expected regret caused by pulling arm $i$ instead of arm
$j$. By comparing (\ref{eq:external}) and (\ref{eq:internal}),
external regret can be bounded above by internal regret as
\cite{Stoltz05}
\begin{equation}
\label{eq:extint}
R_{\textup{Ext}}^{(k)}=\max_{i=1,...,N_{k}}\sum_{j=1}^{N_{k}}R^{(k)}_{(i \to j),n}
\leq N_{k}\max_{i,j= 1,...,N_{k}}R^{(k)}_{(i \to j),n}=N_{k}R^{(k)}_{\textup{Int}}.
\end{equation}
\begin{remark}
\label{re:upperbound}
Throughout the paper, vanishing (zero-average) external and internal
regret means that $\lim_{n \to \infty}\frac{1}{n} R_{\textup{Ext}}=0$
and $\lim_{n \to \infty}\frac{1}{n} R_{\textup{Int}}=0$,
respectively. In other words, we have $R_{\textup{Ext}}\in
o(n)$ and $R_{\textup{Int}}\in o(n)$. Note that by (\ref{eq:extint}),
$R_{\textup{Int}}\in o(n)$ yields $R_{\textup{Ext}}\in o(n)$. Throughout 
the paper, we call any strategy with $R_{\textup{Int}}\in o(n)$ as "no-regret
strategy".
\end{remark}
\subsection{Equilibrium}\label{subsec:Equilib}
From the view point of each player $k$, an MP-MAB is seen as a game
with two agents: player $k$ itself, and the \textit{set} of all other $K-1$
players (referred to as the opponent), whose joint action profile affects the
reward achieved by player $k$. We consider here the most general
framework, where the opponent is non-oblivious, i.e. its series of
actions depends on the actions of player $k$. It is known that a game
against a non-oblivious opponent can be modeled \textit{only} by
adversarial bandit games \cite{Bubeck12}, while similar to other
game-theoretic formulations, the solution is considered to be equilibrium, 
most importantly Nash and correlated equilibria.\footnote{These 
definitions are quite standard (see e.g. \cite{Nisan07}), and thus 
we do not restate them here.}

In the context of game-theoretic bandits, an important result is the 
following theorem.
\begin{theorem}[\cite{Bianchi06}]
\label{th:CorrConvergence}
Consider a $K$-player bandit game, where each player $k$ is provided with an action set of 
cardinality $N_{k}$. Denote the internal regret of player $k$ by $R_{\textup{Int}}^{(k)}$, 
and the set of correlated equilibria by $\mathfrak{C}$. At time $n$, define the empirical 
joint distribution of the game as
\begin{equation}
\label{eq:EmpiricalDistribution}
\hat{\pi}_{n}(\textbf{i})=\frac{1}{n}\sum_{t=1}^{n}\mathbb{I}_{\left \{ \mathbf{I}_{t}=\textbf{i} \right \}},~~\textbf{i}=(i^{(1)},...,i^{(K)})\in \bigotimes_{k=1}^{K}\left \{1,...,N_{K} \right \}.
\end{equation}
Then, if all players $k \in \left \{1,...,K \right \}$ play according to any strategy so that
\begin{equation}
\label{eq:CorrequilibriumTheorem}
\lim_{n \to \infty} \frac{1}{n} R_{\textup{Int}}^{(k)}=0, 
\end{equation}
the distance $\inf _{\pi \in \mathfrak{C}}\sum_{\textbf{i}}\left | \hat{\pi}_{n}(\textbf{i}) -\pi(\textbf{i})\right |$ 
between the empirical joint distribution of plays and the set of correlated equilibria converges 
to $0$ almost surely.
\end{theorem}
Theorem \ref{th:CorrConvergence} simply states that in an MP-MAB game, if all players play 
according to a strategy with vanishing internal regret (no-regret), then the empirical joint 
distribution of plays converges to the set of correlated equilibria. Note that the strategies 
used by players are not required to be identical. Since a rational player is always interested 
in minimizing its regret, the assumption that every player plays according to a no-regret 
strategy is reasonable.
\subsection{From Vanishing External Regret to Vanishing Internal Regret}\label{subsec:Conversion}
In \cite{Stoltz05}, an approach is proposed for converting any selection strategy with 
vanishing external regret to another version with vanishing internal regret. We describe 
this approach briefly. 

Consider a selection strategy (O-strategy, hereafter) which at each time $t$ assigns 
probability distribution $\mathbf{P}_{t}$ to the set of $N$ actions, and selects an 
action according to this distribution. Assume that the player starts using O-Strategy 
with uniform distribution over $N$ actions. At each time $t>1$, the O-strategy has 
already selected $\mathbf{P}_{t-1}=\left (p_{1,t-1},..,p_{i,t-1},..,p_{j,t-1},..,p_{N,t-1}\right)$. 
Now, the O-strategy constructs a meta-strategy (M-strategy, hereafter) with $N(N-1)$ 
virtual strategies based on $\mathbf{P}_{t-1}$. Each virtual strategy corresponds to 
a pair of actions $(i \to j)$, $(i,j \in \left \{1,...,N \right\}, i\neq j)$, and 
constructs a distribution over $N$ actions by assigning the probability mass of action 
$i$ to action $j$. That is, it defines $\mathbf{P}_{t-1}^{(i\to j)}=\left (p_{1,t-1},..,
0,..,p_{j,t-1}+p_{i,t-1},..,p_{N,t-1}\right)$, which has $0$ and $p_{j,t-1}+p_{i,t-1}$ 
at the place of $p_{i,t-1}$ and $p_{j,t-1}$, respectively, and all other elements remain 
unchanged. Assume that the M-strategy treats these virtual strategies as actions. That is, 
at each time $t$, it defines a probability vector $\delta_{t}$ over $N(N-1)$ virtual actions, 
where the probability of action $(i\to j)$, i.e. $\delta_{(i\to j),t}$, depends on its past 
performance.\footnote{Note that the gains of virtual actions cannot be calculated explicitly. 
Later we will see that the gain achieved by any virtual action $(i \to j)$ is calculated 
based on the gain achieved by playing true actions $i$ and $j$.}~Now, at time $t$, the 
O-strategy assigns a distribution $\mathbf{P}_{t}$ to $N$ actions, where $\mathbf{P}_{t}=
\sum_{(i,j):i\neq j}\mathbf{P}_{t}^{(i \to j)}\delta_{(i \to j),t}$. The constructed O-strategy 
has the characteristic that its internal regret is upper-bounded by the external regret of 
the M-strategy over $N(N-1)$ virtual actions according to probability $\delta_{t}$. Thus, 
if the M-strategy exhibits vanishing external regret, the O-strategy results in vanishing 
internal regret. In Section \ref{sec:BEBWA} and \ref{sec:BFPL}, we use this property to 
design no-regret selection strategies. 
\section{Bandit-Theoretical Model of Infrastructureless Wireless Networks}\label{sec:System}
We consider a network consisting of $K$ transmitter-receiver pairs,
denoted by $(k,k')$, where $k,k' \in \left \{1,...,K \right\}$. The 
transmitter-receiver pair $(k,k')$ is referred to as user or player 
$k$. Each user $k$ can access $C_{k}$ mutually orthogonal channels 
at $L_{k}$ quantized power levels. This implies that its strategy 
set includes $N_{k}=C_{k} \times L_{k}$ actions, where at time $t$ 
each action $I_{t}^{(k)}=(c_{t}^{(k)},l_{t}^{(k)})$ consists of one 
channel index (which corresponds to some channel quality), and one 
power level. Therefore, the joint action profile of users, $\mathbf{I}_{t}$, 
is to be understood here as the pair $(\mathbf{c}_{t},\mathbf{l}_{t})$, 
where $\mathbf{c}_{t}=(c^{(1)}_{t},...,c^{(K)}_{t})$ and $\mathbf{l}_{t}=
(l^{(1)}_{t},...,l^{(K)}_{t})$. As each channel might be accessible 
by multiple users, co-channel interference (collision, interchangeably) 
is likely to arise. Since users are allowed to select a new channel 
and to adapt their power levels at each transmission trial, interference 
pattern in general changes over time. In addition, the distribution 
of fading coefficients might be time-varying so that acquiring channel 
and/or network information at the level of autonomous transmitters 
would be extremely challenging and inefficient. Therefore, we assume 
that
\begin{enumerate}[\theassume]
{\refstepcounter{assume}\label{as:NoChannelKnowedge}}
\item transmitters have \textit{no} channel knowledge or any other side 
information such as the number of users or their selected actions.
  {\refstepcounter{assume}\label{as:NoCoordination}}
\item In addition, users do not coordinate their actions that can be chosen 
completely asynchronously by each user.
\end{enumerate}
Note that as users do not observe the actions of each other, it might 
be in their interest to select their actions at the beginning of trials, 
thereby using the remaining time for data transmission.

In this paper, we model the joint channel and power level selection problem 
as a $K$-player adversarial bandit game, where player $k$ decides for one of 
the $N_{k}$ actions. We define the expected utility function (reward) of player 
$k$ to be\footnote{Throughout the paper, logarithms are based 2 unless otherwise 
is stated.}
\begin{equation}
\label{eq:reward}
G^{(k)}_{t}(\mathbf{I}) =\log \biggl(\frac{l^{(k)}|h_{kk',t,c^{(k)}}|^{2}}
{\sum_{q=1}^{Q_{k}} l^{(q)}|h_{qk',t,c^{(k)}}|^{2} + N_{0}}\biggr)-\alpha \cdot l^{(k)}\,,
\end{equation}
for some given joint action profile $\mathbf{I}=(\mathbf{c},\mathbf{l})$. In 
(\ref{eq:reward}), $Q_{k}<K$ is the number of players that interfere with user 
$k$ in channel $c^{(k)}$. Throughout the paper, $|h_{uv,t,c}|^{2} \in \mathbb{R}^{+}$ 
is used to denote the average gain of channel $c$ between $u \to v$ at time $t$. 
$N_{0}$ is the variance of zero-mean additive white Gaussian noise, and $\alpha \geq 0$ 
is the constant power price factor. The last term in (\ref{eq:reward}) is used 
to penalize the use of excessive power. According to Section \ref{sec:Preliminaries}, 
let $g_{t}^{(k)}(\mathbf{I}_{t}) \in [0,1]$ denote the achieved reward of player 
$k$ at time $t$, as a function of joint action profile $\mathbf{I}_{t}$. We consider 
a game with noisy rewards where $g_{t}^{(k)}(\mathbf{I})=G^{(k)}_{t}(\mathbf{I})+
\epsilon_{t}$, with $\epsilon$ being some zero-mean random variable 
with bounded variance, which is independent and identically distributed over time. 
As it is well-known, in a non-cooperative game, the primary goal of each selfish 
player is to maximize its own accumulated reward. Formally, this can be written as
%
\begin{equation}
\label{eq:AggReward}
\textup{maximize}_{(c_{t}^{(k)},l_{t}^{(k)})}~~\sum_{t=1}^{n} g_{t}^{(k)}(\mathbf{c}_{t},\mathbf{l}_{t}),
\end{equation}
where $c_{t}^{(k)} \in \left \{ 1,...,C_{k} \right \} $ and
$l_{t}^{(k)} \in \left \{ 1,...,L_{k} \right \}$. By Assumptions 
\ref{as:NoChannelKnowedge} and \ref{as:NoCoordination}, however, 
it is clear that the objective function in (\ref{eq:AggReward}) 
is not available. For this reason, we argue for a less ambitious 
goal, which is known as \textit{regret minimization}. More precisely, 
each player $k$ attempts to achieve vanishing external regret in 
the sense that
\begin{equation}
\label{eq:RegMin}
\begin{aligned}
 \lim_{n \to \infty} \frac{1}{n}~&R_{\textup{Ext}}^{(k)} \\ 
=\lim_{n \to \infty} & \frac{1}{n}\left (\max_{i=1,...,N_{k}} \sum_{t=1}^{n}g_{t}^{(k)}(i,\mathbf{I}_{t,k}^{-})- 
\sum_{t=1}^{n}\bar{g}_{t}^{(k)}(\mathbf{P}_{t}^{(k)},\mathbf{I}_{t,k}^{-}) \right)=0.
\end{aligned}
\end{equation}
In addition to the individual strategy of each user aiming at satisfying 
(\ref{eq:RegMin}), all players should achieve some steady state, i.e. 
\textit{equilibrium}. Therefore, in the remainder of this paper, we develop 
algorithmic solutions to the resource allocation problem with a twofold 
objective in mind: i) external regret of each user should vanish asymptotically 
according to (\ref{eq:RegMin}) and ii) the actions of all players should 
convergence to equilibrium.

By (\ref{eq:extint}), the external regret of each user is upper-bounded by 
its internal regret. As a result, if all users select their actions according to some 
no-regret strategy, not only (\ref{eq:RegMin}) is achieved by all of them 
(see also Remark \ref{re:upperbound}), but also the corresponding game converges 
to equilibrium in some sense, which immediately follows from Theorem 
\ref{th:CorrConvergence}. In Sections \ref{sec:BEBWA} and \ref{sec:BFPL}, 
we present two internal-regret minimizing strategies that are shown to solve 
the game and, with it, to achieve the two objectives mentioned above. Both 
algorithms can be applied in a \textit{fully decentralized} manner by each player, 
since at each time, they only require the set of past rewards of the respective 
player.

Finally, it is worth noting that the set of correlated equilibria for the general 
time-varying repeated game defined by (\ref{eq:reward}) cannot be characterized. 
Nevertheless, in what follows, we characterize this set for two games defined by 
some relaxed versions of (\ref{eq:reward}). First, consider a game similar to the 
one defined above, with the difference that unlike (\ref{eq:reward}), the reward 
process is assumed to be stationary, i.e.
\begin{equation}
\label{eq:rewardT}
G^{(k)}(\mathbf{I}) =\log \biggl(\frac{l^{(k)}|h_{kk',c^{(k)}}|^{2}}
{\sum_{q=1}^{Q_{k}} l^{(q)}|h_{qk',c^{(k)}}|^{2} + N_{0}}\biggr)-\alpha \cdot l^{(k)}\,,
\end{equation}
which implies that the average channel gains are time-invariant. By the following proposition, 
this game has a unique correlated equilibrium. 
\begin{proposition}
\label{pr:EqType}
Consider a $K$-player game where the expected reward function of each player $k$ is defined 
by (\ref{eq:rewardT}). This game has a unique correlated equilibrium which places probability 
one on its unique pure-strategy Nash equilibrium.
\end{proposition}
\begin{IEEEproof}
See Section \ref{subsec:CorEq-Pr}.
\end{IEEEproof}
%

Now let the expected reward function be defined as follows:
\begin{equation}
\label{eq:rewardTwo}
G^{(k)}(\mathbf{I}) = \log\bigl(l^{(k)}\frac{|h_{kk',c^{(k)}}|^{2}}{N_{0}}\bigr)-\alpha l^{(k)},
\end{equation}
which is more restricted, but simpler than (\ref{eq:rewardT}). With this choice of expected 
reward function, the game can be shown to have a unique correlated equilibrium that maximizes 
the aggregate utility of all players, i.e. the social welfare. This result is stated formally 
in the following proposition.
\begin{proposition}
\label{pr:EqTypeTwo}
Consider a $K$-player game where each player $k$ has the expected reward function $G^{(k)}(\cdot)$ 
given by (\ref{eq:rewardTwo}). This game has a unique correlated equilibrium which places probability 
one on a unique pure strategy Nash equilibrium that maximizes $\sum_{k=1}^{K}G^{(k)}(\cdot)$.
\end{proposition}
\begin{IEEEproof}
See Section \ref{subsec:CorEq-Pr-Two}.
\end{IEEEproof}
\section{No-Regret Bandit Exponential-Based Weighted Average Strategy}\label{sec:BEBWA}
The basic idea of an exponential-based weighted strategy is to assign
each action, at every trial, some selection probability which is
inversely proportional to exponentially-weighted accumulated regret
(or directly proportional to exponentially-weighted accumulated
reward) caused by that action in the past \cite{Hart01}. Roughly
speaking, if playing an action has resulted in large regret in the
past, its future selection probability is small, and vice versa.

As described in Section \ref{subsec:Regret}, in bandit formulation,
players only observe the reward of the played action, and not those of
others. Therefore the reward of each action $i$ is estimated as
\cite{Bianchi06}
\begin{equation} 
\label{eq:GainEstim}
\tilde{g}_{t}^{(k)}(i)=\left\{\begin{matrix}
\frac{g_{t}^{(k)}(I_{t}^{(k)})}{p_{i,t}^{(k)}} & i=I_{t}^{(k)}\\ 
0 & o.w.
\end{matrix},\right.
\end{equation}
which is an unbiased estimate of the true reward of action $i$; 
that is, $\textup{E}_{t}\left [\tilde{g}^{(k)}(i)\right ]=g^{(k)}(i)$. Estimated rewards 
are afterwards used to calculate regrets. For example, the regret of \textit{not} playing action 
$j$ instead of action $i$ yields 
\begin{equation} 
\label{eq:deltaAccRegret}
\tilde{R}^{(k)}_{(i \to j),t-1}=\sum_{s=1}^{t-1}\tilde{r}^{(k)}_{(i \to j),s}=\sum_{s=1}^{t-1}
p_{i,s}^{(k)}(\tilde{g}_{s}^{(k)}(j)-\tilde{g}_{s}^{(k)}(i)).
\end{equation}

Despite exhibiting vanishing external regret, weighted average
strategies yield in general large internal regret; as a result, even
if all players play according to such strategies, the game does
\textit{not} converge to equilibrium. In the following, we utilize the
bandit version of exponentially weighted average strategy
\cite{Bianchi03}, and convert it to an improved version that yields
small \textit{internal} regret, using the approach of Section
\ref{subsec:Conversion}. The strategy is called no regret bandit
exponentially-weighted average strategy (NR-BEWAS), and is described
in Algorithm \ref{alg:NR-BEWAS}.
\begin{algorithm}
\caption{No-Regret Bandit Exponential-Based Weighted Average Strategy (NR-BEWAS)}
\label{alg:NR-BEWAS}
\small
\begin{algorithmic}[1]
\STATE If the game horizon, $n$, is known, define $\gamma_{t}$ and $\eta_{t}$ as given in 
Proposition \ref{pr:NR-BEWAS-One}, otherwise as those given in Proposition \ref{pr:NR-BEWAS}.  
\STATE Define $\Phi (\textbf{U})=\frac{1}{\eta_{t}}\ln\left (\sum_{i=1}^{N_{k}} \exp(\eta _{t}u_{i})\right)$, 
       where $\textbf{U}=(u_{1},...,u_{N_{k}})\in \mathbb{R}^{N_{k}}$.
\STATE Let $\mathbf{P}_{1}^{(k)}=\left (\frac{1}{N_{k}},...,\frac{1}{N_{k}}\right)$ 
              (uniform distribution).
\STATE Select an action using $\mathbf{P}_{1}^{(k)}$.
\STATE Play and observe the reward.
  \FOR {$t=2,...,n$} 
     \STATE Let $\mathbf{P}_{t-1}^{(k)}$ be the mixed strategy 
            at time $t-1$, i.e. $\mathbf{P}_{t-1}^{(k)}=\left                                                                                                  (p_{1,t-1}^{(k)},..,p_{i,t-1}^{(k)},..,p_{j,t-1}^{(k)},..,p_{N_{k},t-1}^{(k)}\right)$.
     \STATE Construct $\mathbf{P}_{t-1}^{{(k)},(i \to j)}$ as follows: replace 
            $p_{i,t-1}^{(k)}$ in $\mathbf{P}_{t-1}^{(k)}$ by zero, and instead 
            increase $p_{j,t-1}^{(k)}$ to $p_{j,t-1}^{(k)}+p_{i,t-1}^{(k)}$. Other 
            elements remain unchanged. We obtain $\mathbf{P}_{t-1}^{{(k)},(i\to                                                                                                j)}=\left(p_{1,t-1}^{(k)},..,0,..,p_{j,t-1}^{(k)}+p_{i,t-1}^{(k)},..,p_{N_{k},t-1}^{(k)}\right )$.         
     \STATE Define 
            \begin{equation}
            \label{eq:delta}
            \delta_{(i \to j),t}^{(k)}=\frac{\exp\left ( \eta_{t} \tilde{R}_{(i \to j),t-1}^{(k)}\right)}
            {\sum_{(m \to l):m\neq l}\exp\left ( \eta_{t} \tilde{R}_{(m \to l),t-1}^{(k)}\right )},
            \end{equation}
            where $\tilde{R}_{(i \to j),t-1}^{(k)}$ is calculated by using (\ref{eq:GainEstim}) and (\ref{eq:deltaAccRegret}). 
    \STATE  Given $\delta_{(i \to j),t}^{(k)}$, solve the following fixed point equation to find $\mathbf{P}_{t}^{(k)}$:
            \begin{equation}
            \label{eq:deltaProbA} 
            \mathbf{P}_{t}^{(k)}=\sum_{(i \to j):i\neq j}\mathbf{P}_{t}^{{(k)},(i \to j)}\delta_{(i \to j),t}^{(k)}.
            \end{equation}
    \STATE  Final probability distribution yields
            \begin{equation}
            \label{eq:deltaProbFinalA}
            \mathbf{P}_{t}^{(k)}=(1-\gamma _{t})\mathbf{P}_{t}^{(k)}+\frac{\gamma _{t}}{N_{k}}.
            \end{equation}
    \STATE  Using the final $\mathbf{P}_{t}^{(k)}$, given by (\ref{eq:deltaProbFinalA}), 
            select an action. 
    \STATE  Play and observe the reward.
    \ENDFOR
\end{algorithmic}
\end{algorithm}

From Algorithm \ref{alg:NR-BEWAS}, NR-BEWAS has two parameters, namely $\gamma_{t}$ and $\eta_{t}$. 
In the event that the game horizon, $n$, is known in advance, these two parameters are constant over 
time ($\eta_{t}=\eta$ and $\gamma_{t}=\gamma$), and the growth rate of regret can be bounded precisely, 
mainly based on the results of \cite{Bianchi06}. Otherwise, they vary with time. In this case, vanishing 
(sub-linear in time) internal regret can be guaranteed; nevertheless, this bound might be loose. This 
discussion is formalized by following propositions.
\begin{proposition}
\label{pr:NR-BEWAS-One}
Let $\eta_{t}=\eta=\left (\frac{\ln N_{k}}{2N_{k}n} \right
)^{\frac{2}{3}}$ and $\gamma_{t}=\gamma=\left (\frac{N_{k}^{2}\ln 
N_{k}}{4n} \right )^{\frac{1}{3}}$. Then Algorithm\ref{alg:NR-BEWAS} 
(NR-BEWAS) yields vanishing internal regret and we have $R^{(k)}_{\textup{Int}}
\in O((nN_{k}^{2}\ln N_{k})^{\frac{2}{3}})$.
\end{proposition} 
\begin{IEEEproof}
See Appendix \ref{subsec:NR-BEWAS-Pr-One}.
\end{IEEEproof}
\begin{proposition}
\label{pr:NR-BEWAS}
Let $\eta_{t}=\frac{\gamma_{t}^{3}}{N_{k}^{2}}$ and
$\gamma_{t}=t^{-\frac{1}{3}}$. Then Algorithm \ref{alg:NR-BEWAS}
(NR-BEWAS) yields vanishing internal regret; that is we have
$R^{(k)}_{\textup{Int}}\in o(n)$.
\end{proposition} 
\begin{IEEEproof}
See Appendix \ref{subsec:NR-BEWAS-Pr}.
\end{IEEEproof}

The following corollaries follow from the above propositions 
and Theorem \ref{th:CorrConvergence}.
\begin{corollary}
\label{cr:NR-BEWAS}
If all players play according to NR-BEWAS, then the empirical joint
frequencies of play converge to the set of correlated equilibria.
\end{corollary}
\begin{IEEEproof}
The proof is a direct consequence of Theorem \ref{th:CorrConvergence} 
and Proposition \ref{pr:NR-BEWAS-One} or Proposition \ref{pr:NR-BEWAS}.
\end{IEEEproof}
\begin{corollary}
\label{cr:NR-BEWAS-Rate}
Let $\epsilon$-correlated equilibrium approximate correlated
equilibrium in the sense that $\bigcap_{\epsilon>0}
\mathfrak{C}_{\epsilon}=\mathfrak{C}$. Assuming that the game horizon
is known and all players play according to NR-BEWAS, then the minimum
required number of trials to achieve $\epsilon$-correlated equilibrium
yields $\max_{k=1,...,K}\epsilon^{-\frac{3}{2}} O\left ((N_{k}K)(N_{k}^{2}
\ln N_{k}+K^{2}\ln K) \right )$, which is proportional to $\epsilon^{-\frac{3}{2}}$ 
and increases polynomially in the number of actions as well as in the 
number of players.
\end{corollary}
\begin{IEEEproof}
The proof follows from the bound of Proposition \ref{pr:NR-BEWAS-One} 
and Remark 7.6 of \cite{Bianchi06}.\footnote{Details are omitted to 
avoid unnecessary restatement of existing analysis.}
\end{IEEEproof}
\section{No-Regret Bandit Follow the Perturbed Leader Strategy}\label{sec:BFPL}
Similar to the weighted-average strategy presented in the previous section, 
the strategy \textit{follow the perturbed leader} is an approach to solve 
online decision-making problems. In the basic version of this approach, 
called \textit{follow the leader} \cite{hannan57}, the action with the 
minimum regret in the past is selected at each trial. However, this method 
is deterministic and therefore does not achieve vanishing regret against 
non-oblivious opponents. Therefore, in \textit{follow the perturbed leader}, 
player adds a random perturbation to the vector of accumulated regrets, and 
the action with the minimum perturbed regret in the past is selected
\cite{Bianchi06}. In \cite{Kujala07}, a bandit version of this algorithm 
is constructed, where unobserved rewards are estimated. The authors show 
that the developed algorithm exhibits vanishing \textit{external regret}. 
Similar to NR-BEWAS, we here modify the algorithm of \cite{Kujala07} to 
ensure vanishing \textit{internal regret}. The approach is called no-regret 
bandit follow the perturbed leader strategy (NR-BFPLS).

\begin{algorithm}
\caption{No-Regret Bandit Follow the Perturbed Leader Strategy (NR-BFPLS)}
\label{alg:NR-BFPLS}
\small
\begin{algorithmic}[1]
\STATE Define $\epsilon_{t}=\epsilon_{n}=\frac{\sqrt{\ln n}}{3\sqrt{N_{k}n}}$, 
       and $\gamma_{t}=\min(1,N_{k}\epsilon_{t})$. Note that unlike  
       NR-BEWAS, here we know the game horizon ($n$) in advance.
\STATE Let $\mathbf{P}_{1}^{(k)}=\left (\frac{1}{N_{k}},...,\frac{1}{N_{k}}\right)$ 
       (uniform distribution).
\STATE Select an action using $\mathbf{P}_{1}^{(k)}$.
\STATE Play and observe the reward.
  \FOR {$t=2,...,n$} 
    \STATE Let $\mathbf{P}_{t-1}^{(k)}$ be the mixed 
           strategy at time $t-1$, i.e. $\mathbf{P}_{t-1}^{(k)}=\left                                                                                   (p_{1,t-1}^{(k)},..,p_{i,t-1}^{(k)},..,p_{j,t-1}^{(k)},..,p_{N_{k},t-1}^{(k)}\right)$.
    \STATE Construct $\mathbf{P}_{t-1}^{{(k)},(i \to j)}$ as follows: replace 
           $p_{i,t-1}^{(k)}$ in $\mathbf{P}_{t-1}^{(k)}$ by zero, and instead 
           increase $p_{j,t-1}^{(k)}$ to $p_{j,t-1}^{(k)}+p_{i,t-1}^{(k)}$. Other 
           elements remain unchanged. We obtain $\mathbf{P}_{t-1}^{{(k)},(i\to                                                                                                j)}=\left(p_{1,t-1}^{(k)},..,0,..,p_{j,t-1}^{(k)}+p_{i,t-1}^{(k)},..,p_{N_{k},t-1}^{(k)}\right )$.         
    \STATE Calculate $\tilde{R}_{(i \to j),t-1}^{(k)}$ using (\ref{eq:GainEstim}) and (\ref{eq:deltaAccRegret}). 
    \STATE Define $\sigma_{(i \to j),t-1}=\left(\sum_{\tau=1}^{t-1}\frac{1}{\delta_{(i \to j),t}^{(k)}}\right )^{\frac{1}{2}}$, 
           which is the upper-bound of conditional variances of random variables $\tilde{R}_{(i \to j),t-1}^{(k)}$ \cite{Kujala07}.
    \STATE Let $\tilde{R}_{(i \to j),t-1}^{(k)}=\tilde{R}_{(i \to j),t-1}^{(k)}- 
           \sqrt{1+\sqrt{2/N_{k}}}~\sigma_{(i \to j),t-1}~\sqrt{\ln(t)}$ \cite{Kujala07}.              
    \STATE Randomly select a perturbation vector $\underline{\mu_{t}}$ with $N_{k}(N_{k}-1)$ 
           elements from two-sided exponential distribution with width $\epsilon_{t}$.
    \STATE Consider a selection rule which selects the action $(i \to j)$ given by
           \begin{equation}
           \label{eq:SelAct}
           \textup{argmax}\left \{\tilde{R}_{(i \to j),t-1}^{(k)}+\mu_{(i \to j),t} \right \},
           ~~(i \to j)\in \left \{1,...,N_{k}(N_{k}-1)\right \}
           \end{equation}
           Note that in our setting $\tilde{R}_{(i \to j)}$ denotes the estimated regret 
           of \textit{not} playing action $(i \to j)$, hence we find the action 
           with largest $\tilde{R}$.
    \STATE From (\ref{eq:SelAct}), calculate the probability $\delta_{(i \to j),t}^{(k)}$ 
           assigned to each pair $(i \to j)$.    
    \STATE Given $\delta_{(i \to j),t}^{(k)}$, solve the following fixed point 
           equation to find $\mathbf{P}_{t}^{(k)}$.
           \begin{equation}
           \label{eq:deltaProb} 
           \mathbf{P}_{t}^{(k)}=\sum_{(i \to j):i\neq j}\mathbf{P}_{t}^{{(k)},(i \to j)}\delta_{(i \to j),t}^{(k)}.
           \end{equation}
    \STATE Final probability distribution yields
           \begin{equation}
           \label{eq:deltaProbFinal}
           \mathbf{P}_{t}^{(k)}=(1-\gamma_{t})\mathbf{P}_{t}^{(k)}+\frac{\gamma_{t}}{N_{k}}.
           \end{equation}
    \STATE Using the final $\mathbf{P}_{t}^{(k)}$, given by (\ref{eq:deltaProbFinal}), 
           select an action. 
    \STATE Play and observe the reward.
\ENDFOR
\end{algorithmic}
\end{algorithm}
Algorithm \ref{alg:NR-BFPLS} requires the knowledge of the probability
assigned to each action by the \textit{follow the perturbed
  leader} strategy at every trial. However, in contrast to NR-BEWAS,
these probabilities are not assigned explicitly; therefore we explain
how to calculate these values.\\
From (\ref{eq:SelAct}), the selection probability of virtual action
$(i \to j)\in \left \{ 1,...,N_{k}(N_{k}-1)\right \}$ is the
probability that $\tilde{R}_{(i \to j),t-1}$ plus perturbation
$\mu_{(i \to j),t}$ is larger than those of other actions, i.e.
\begin{equation}
\label{eq:ProbCal}
\begin{aligned}
\Pr&[I_{t}=(i \to j)]\\
= &\Pr[\tilde{R}_{(i \to j),t-1}+\mu_{(i \to j),t}\geq \tilde{R}_{(i' \to j'),t-1}+
\mu_{(i' \to j'),t} ~\forall (i \to j)\neq (i' \to j')] \\ 
 = & \int_{-\infty}^{\infty}\Pr[\tilde{R}_{(i \to j),t-1}+\mu_{(i \to j),t}=
 m \wedge \tilde{R}_{(i' \to j'),t-1}+\mu_{(i' \to j'),t}\leq m ~\forall (i \to j)\neq (i' \to j')]dm\\ 
 =&\int_{-\infty}^{\infty} \Pr[\tilde{R}_{(i \to j),t-1}+\mu_{(i \to j),t}=m]
 \prod_{(i' \to j')\neq (i \to j)}\Pr[\tilde{R}_{(i' \to j'),t-1}+\mu_{(i' \to j'),t}\leq m] dm.\\
\end{aligned}
\end{equation}
Since $\mu_{t}$ is distributed according to a two-sided exponential distribution 
with width $\epsilon_{n}$, the terms under integral can be calculated easily 
(see \cite{Hutter05}, for example).
Now we are in a position to show some properties of NR-BFPLS (Algorithm \ref{alg:NR-BFPLS}).
\begin{proposition}
\label{pr:NR-BFPL}
Let $\epsilon_{t}=\epsilon=\frac{\sqrt{\ln n}}{3\sqrt{N_{k}n}}$ and
$\gamma_{t}=\gamma=\min(1,N_{k}\epsilon_{t})$. Then Algorithm
\ref{alg:NR-BFPLS} (NR-BFPL) yields vanishing internal
regret with $R^{(k)}_{\textup{Int}}\in
O((nN_{k}^{2}\ln N_{k})^{\frac{1}{2}})$.
\end{proposition} 
\begin{IEEEproof}
By \cite{Kujala07}, we know that if the BPFL algorithm is applied to $N_{k}$ actions, 
then $R^{(k)}_{\textup{Ext}}\in O((nN_{k}\ln N_{k})^{\frac{1}{2}})$. Using this, the 
proof proceeds along similar lines as the proof of Proposition \ref{pr:NR-BEWAS-One} 
and is therefore omitted here.
\end{IEEEproof}

\begin{corollary}
\label{cr:NR-BFPL-Rate}
Assuming that the game horizon is known and all players play according
to NR-BFPLS, then the minimum required number of trials to achieve
$\epsilon$-correlated equilibrium yields $\max_{k=1,...,K}\epsilon^{-2} 
O\left ((N_{k}K)(N_{k}^{2}\ln N_{k}+K^{2}\ln K)\right )$, which is 
proportional to $\epsilon^{-2}$ and increases polynomially in the 
number of actions as well as in the number of players.
\end{corollary}
\begin{IEEEproof} 
The proof is a result of the bound of Proposition \ref{pr:NR-BFPL} and 
Remark 7.6 of \cite{Bianchi06}.
\end{IEEEproof}

%
%
\section{Bandit Experimental Regret-Testing Strategy}\label{sec:BERT}
Experimental regret-testing belongs to the large family of exhaustive search algorithms, and is 
comprehensively discussed in \cite{Germano07} and \cite{Bianchi06} for bandit games. In this
section, we briefly review this approach, and investigate its performance later in Section 
\ref{subsec:NumericOne}. 

First, the time is divided into periods $m=1,2,...$ of length $T$ so that for each $m$ we have 
$t \in [(m-1)T+1,mT]$. At the beginning of period $m$, any player $k$ randomly selects a mixed 
strategy, denoted by $\textup{P}_{m}^{(k)}$. Moreover, some random variable $U_{k,t}^{(m)} \in 
\left \{1,...,n_{k},...,N_{k} \right \}$ is defined as follows. For $t \in [(m-1)T+1,mT]$, and 
for each $n_{k}$, there are exactly $s$ values of $t$ such that $U_{k,t}^{(m)}=n_{k}$, and $U_{k,t}^{(m)}=0$ 
for the remaining $t=T-sN_{k}$ trials. At time $t$, the action $I_{t}^{(k)}$ is selected to be 
\cite{Bianchi03}
\begin{equation}
I_{t}^{(k)}:\left\{\begin{matrix}
\textup{is distributed as}~\textup{P}_{m}^{(k)}& \textup{if}~~U_{k,t}^{(m)}=0\\ 
\textup{equals}~n_{k}& ~~\textup{if}~~U_{k,t}^{(m)}=n_{k}
\end{matrix}\right..
\end{equation}
At the end of period $m$, player $k$ calculates the experimental regret of playing each action 
$n_{k}$ as \cite{Bianchi03}
\begin{equation}
\label{eq:exprimental}
\hat{r}_{m,n_{k}}^{(k)}=\frac{1}{T-sN_{k}}\sum_{t=(m-1)T+1}^{mT}g_{t}^{(k)}(\mathbf{I}_{t})
\mathbb{I}_{\left \{ U_{k,t}^{(m)} =0\right \}} -\frac{1}{s}\sum_{t=(m-1)T+1}^{mT}g_{t}^{(k)}
(n_{k},\mathbf{I}_{t,k}^{-})\mathbb{I}_{\left \{ U_{k,t}^{(m)}=n_{k}\right \}}. 
\end{equation}
If the regret is smaller than an acceptable threshold $\rho$, the player continues to play its 
current mixed strategy. Otherwise, another mixed strategy is selected. The procedure is summarized 
in Algorithm \ref{alg:BERTS}. It is known that if the parameters of BERTS (e.g. $T$ and $\rho$) 
are chosen appropriately, then, in a long run, the played mixed strategy profile is an approximate 
Nash equilibrium for almost all the time. Details can be found in \cite{Bianchi06}, and hence are 
omitted.
\begin{algorithm}
\caption{Bandit Experimental Regret Testing Strategy \cite{Bianchi06} (BERTS)}
\label{alg:BERTS}
\small
\begin{algorithmic}[1]
\STATE Set $T$ (period length), $\rho$ (acceptable regret threshold), $\xi \ll 1$ 
(exploration parameter), $m=1$ (period index). Notice that for each period $m=1,...,M$, 
we have $t \in [(m-1)T+1,mT]$.
\STATE Select a mixed strategy, $\textup{P}_{m}^{(k)}$ according to the uniform 
       distribution, from the probability simplex with $N_{k}$ dimensions.
\STATE For each $n_{k} \in \left \{ 1,..,N_{k} \right \}$ select $s$ exploring 
       trials at random. Exploration trials which are dedicated to different actions should 
       not overlap. 
       \FOR{$t=(m-1)T+y$, where $1\leq y<T$}
             \IF {$t$ is an exploring trial dedicated to action $i$}
                 \STATE play action $i$ and observe the reward.
             \ELSE
                 \STATE select an action using $\textup{P}_{m}^{(k)}$. Play and 
                        observe the reward.
             \ENDIF
       \ENDFOR
       \STATE Calculate the experimental regret of period $m$, $\hat{r}_{m,n_{k}}^{(k)}$, using (\ref{eq:exprimental});
              \IF {
                $\underset{n_{k}=1,...,N_{k}}{\max}~~\hat{r}_{m,n_{k}}^{(k)}>\rho$,                                  
                }
                  \STATE 1) set $m=m+1$, 2) go to line 2.
              \ELSE
                  \STATE
                  \begin{itemize}
                  \item with probability $\xi $: 1) set $m=m+1$, 2) go to line 2;
                  \item with probability $1-\xi$: 1) let $\textup{P}_{m+1}^{(k)}=\textup{P}_{m}^{(k)}$, 
                        2) set $m=m+1$, 3) go to line 3.
                  \end{itemize}
             \ENDIF      
\end{algorithmic}
\end{algorithm}
\section{Numerical Analysis}\label{subsec:Numerical}
Numerical analysis consists of two parts. In Section \ref{subsec:NumericOne}, 
we consider a simple network, and clarify the work flow of algorithms. In 
Section \ref{subsec:NumericTwo}, we consider a larger network, and study 
the performance of the proposed game model and algorithmic solutions in 
comparison with some other selection strategies.
\subsection{Part One}\label{subsec:NumericOne}
\subsubsection{Network model}\label{subsec:SimScenario}
The network consists of two transmitter-receiver pairs (users). There exist two orthogonal 
channels, $C_{1}$ and $C_{2}$, and two power-levels, $P_{1}$ and $P_{2}$. Hence, the action 
set of each user yields $\left\{a_{1}:(C_{1},P_{1}),a_{2}:(C_{1},P_{2}), a_{3}:(C_{2},P_{1}), 
a_{4}:(C_{2},P_{2})\right\}$. The distribution of channel gains changes at each trial. We 
assume that the variance of mean values of these distributions is relatively small, which 
corresponds to \textit{low dynamicity}.\footnote{Note that this assumption is made in order 
to simplify the implementation; as established theoretically, all proposed procedures converge 
to equilibrium for arbitrary varying distributions.}~Channel matrices are 
$H_{1}=\begin{bmatrix}
\left [0.50,0.80 \right] & \left [0.15,0.20 \right]\\ 
\left [0.01,0.05 \right]& \left [0.01,0.09 \right]
\end{bmatrix}$ 
and 
$H_{2}=\begin{bmatrix}
\left [0.02,0.05 \right] & \left [0.02,0.06 \right]\\ 
\left [0.05,0.15 \right]& \left [0.75,0.95 \right]
\end{bmatrix}$,
where $H_{l,(u,v)}$ ($u,v,l \in \left \{1,2 \right \}$), corresponds to the link 
$u \to v$ through channel $l$, and presents the interval from which the mean value 
of the distribution of channel gain is selected at each trial. Moreover, we assume 
$P_{1}=1$, $P_{2}=5$ and $\alpha=10^{-3}$. Except for their instantaneous rewards, 
no other information is revealed to users. This information can be provided by 
the receiver feedback to transmitter. With these settings, it is easy to see that 
$((C_{1},P_{2}),(C_{2},P_{2}))$ is the unique pure strategy Nash equilibrium of 
this game, i.e. the theoretical convergence point. 

\subsubsection{Results and Discussion}\label{subsec:Discussion}
\begin{figure}[t]
\centering
\includegraphics[width=0.7\textwidth]{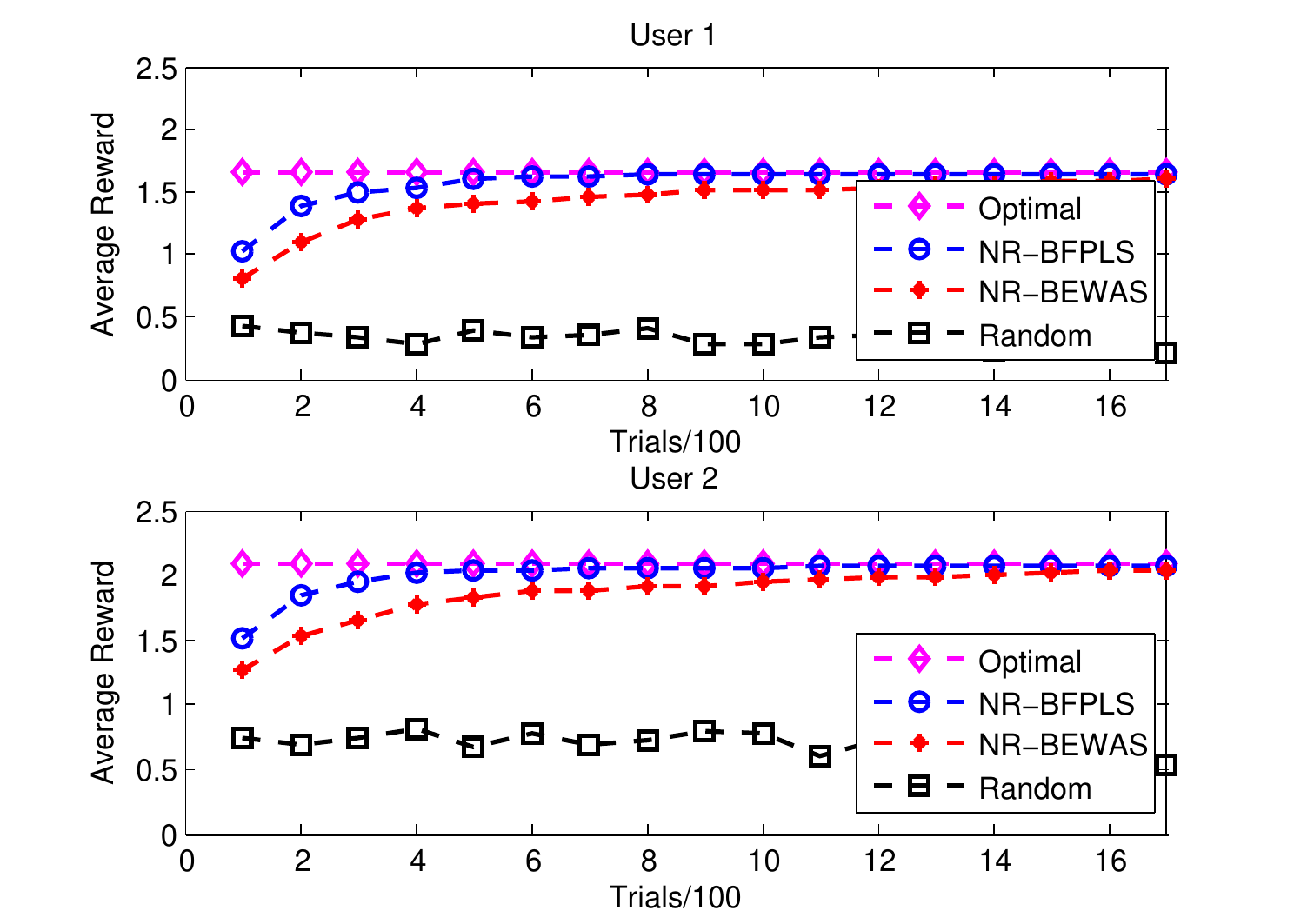}
\caption{Performance of four selection strategies. Both NR-BEWAS and NR-BFPLS exhibit 
vanishing regret; that is, their average rewards converge to that of optimal (centralized) 
selection.}
\label{Tg}
\end{figure}

We investigate the performance of selection strategies NR-BEWAS, NR-BFPLS and BERTS. The 
following strategies are also considered as benchmark:
\begin{itemize}
\item optimal (centralized) action (channel and power level) assignment that is based on 
global statistical channel knowledge and is performed by a central unit. 
\item uniformly random selection.
\end{itemize}
Figure \ref{Tg} compares the average reward achieved by NR-BEWAS and NR-BFPLS by those of 
random and optimal selections. 
From the figure, despite being provided with only strictly limited information, both 
NR-BFPLS and NR-BEWAS exhibit vanishing regret, in the sense that the achieved average 
reward converges to that of centralized scenario. 

\begin{figure}[t]
\centering
\includegraphics[width=0.45\textwidth]{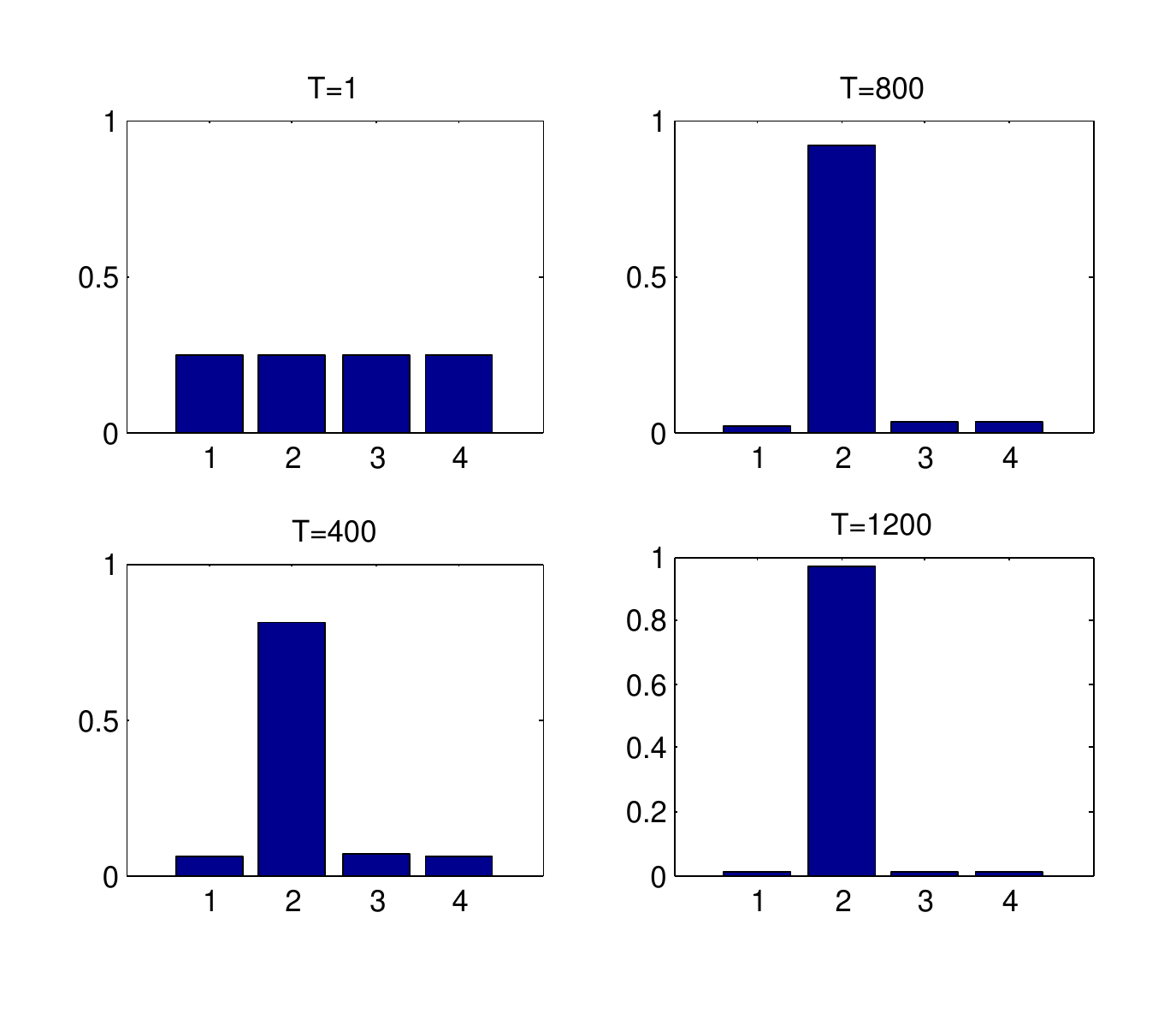}
\caption{Evolution of the mixed strategy of User 1, applying NR-BEWAS. Horizontal axis 
denotes the action indices, where index $i$, $i \in \left \{1,2,3,4 \right\}$, stands 
for action $a_{i}$. Vertical axis shows the weight of each action in the mixed strategy, 
i.e. its probability of being selected. The mixed strategy of User 1 converges to 
$\pi_{1}=(0,1,0,0)$.}
\label{BEWASO}
\end{figure}

\begin{figure}[t]
\centering
\includegraphics[width=0.45\textwidth]{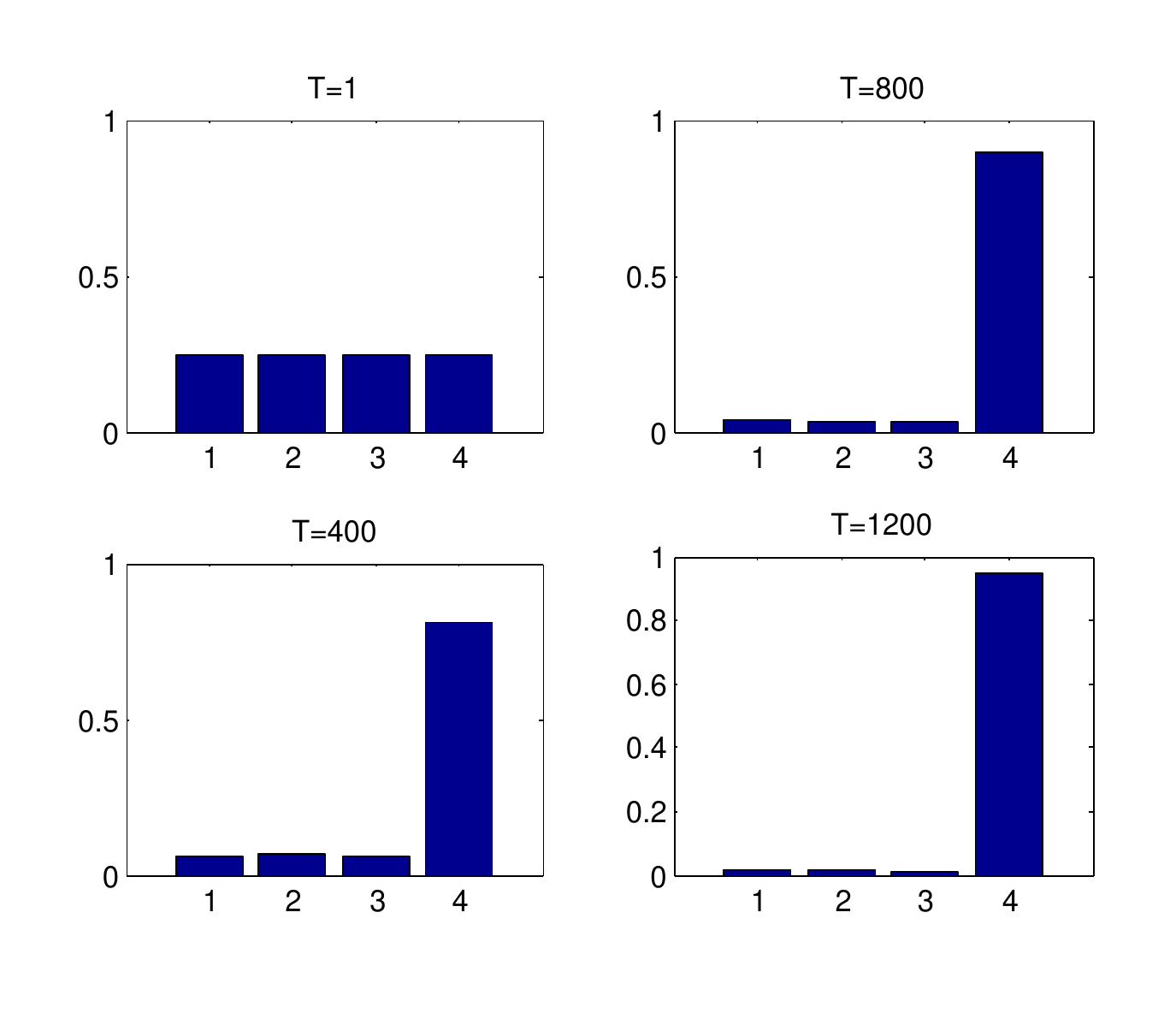}
\caption{Evolution of the mixed strategy of User 2, applying NR-BEWAS. The horizontal 
and vertical axes respectively depict the indices of actions and their selection probabilities. 
The mixed strategy of User 2 converges to $\pi_{2}=(0,0,0,1)$.}
\label{BEWAST}
\end{figure}

\begin{figure}[t]
\centering
\includegraphics[width=0.45\textwidth]{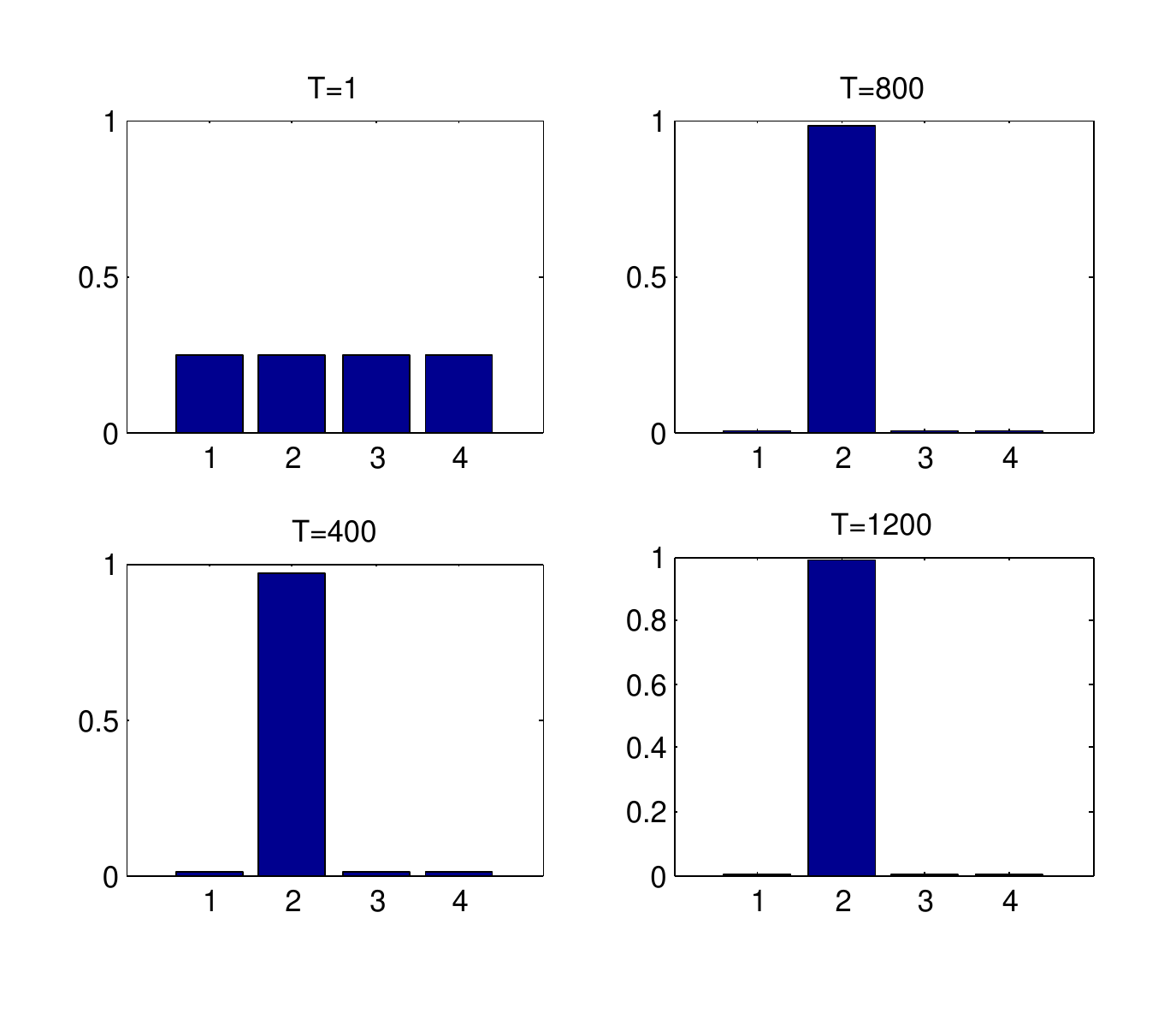}
\caption{Evolution of the mixed strategy of User 1, applying NR-BFPLS. The horizontal and 
vertical axes respectively depict the indices of actions and their selection probabilities. 
The mixed strategy of User 1 converges to $\pi_{1}=(0,1,0,0)$. }
\label{BFPLSO}
\end{figure}

\begin{figure}[t]
\centering
\includegraphics[width=0.45\textwidth]{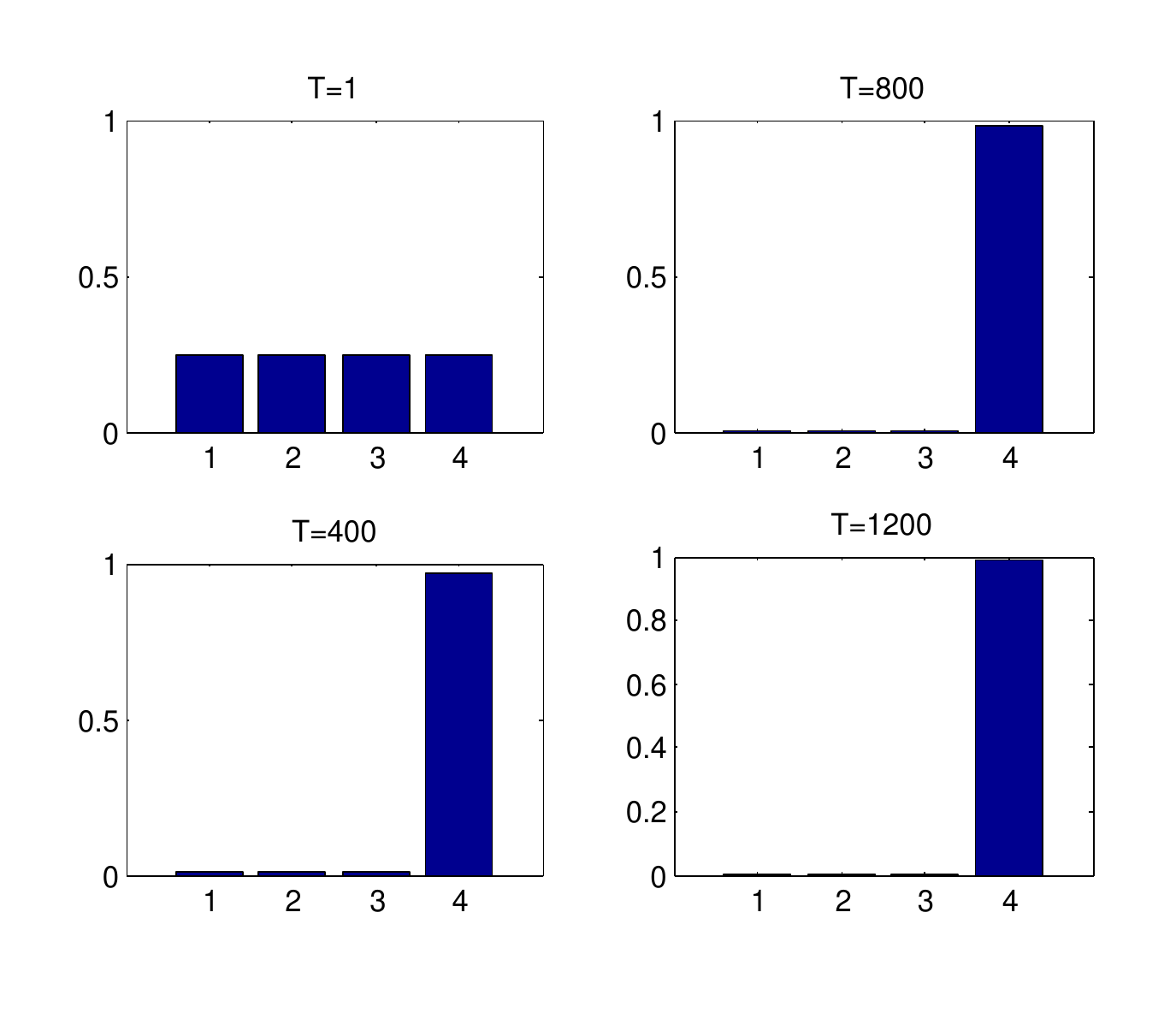}
\caption{Evolution of the mixed strategy of User 2, applying NR-BFPLS. The horizontal and 
vertical axes respectively depict the indices of actions and their selection probabilities. 
The mixed strategy of User 2 converges to $\pi_{2}=(0,0,0,1)$. }
\label{BFPLST}
\end{figure}

Figures \ref{BEWASO} and \ref{BEWAST} illustrate the evolution of mixed strategies of 
the two users when NR-BEWAS is used. Figures \ref{BFPLSO} and \ref{BFPLST}, on the other 
hand, show the same variable when actions are selected by using NR-BFPLS. For both cases, 
the first and second users respectively converge to $a_{2}:(C_{1},P_{2})$ and $a_{4}:(C_{2},P_{2})$, 
as suggested by the theory.

\begin{figure}[t]
\centering
\includegraphics[width=0.98\textwidth]{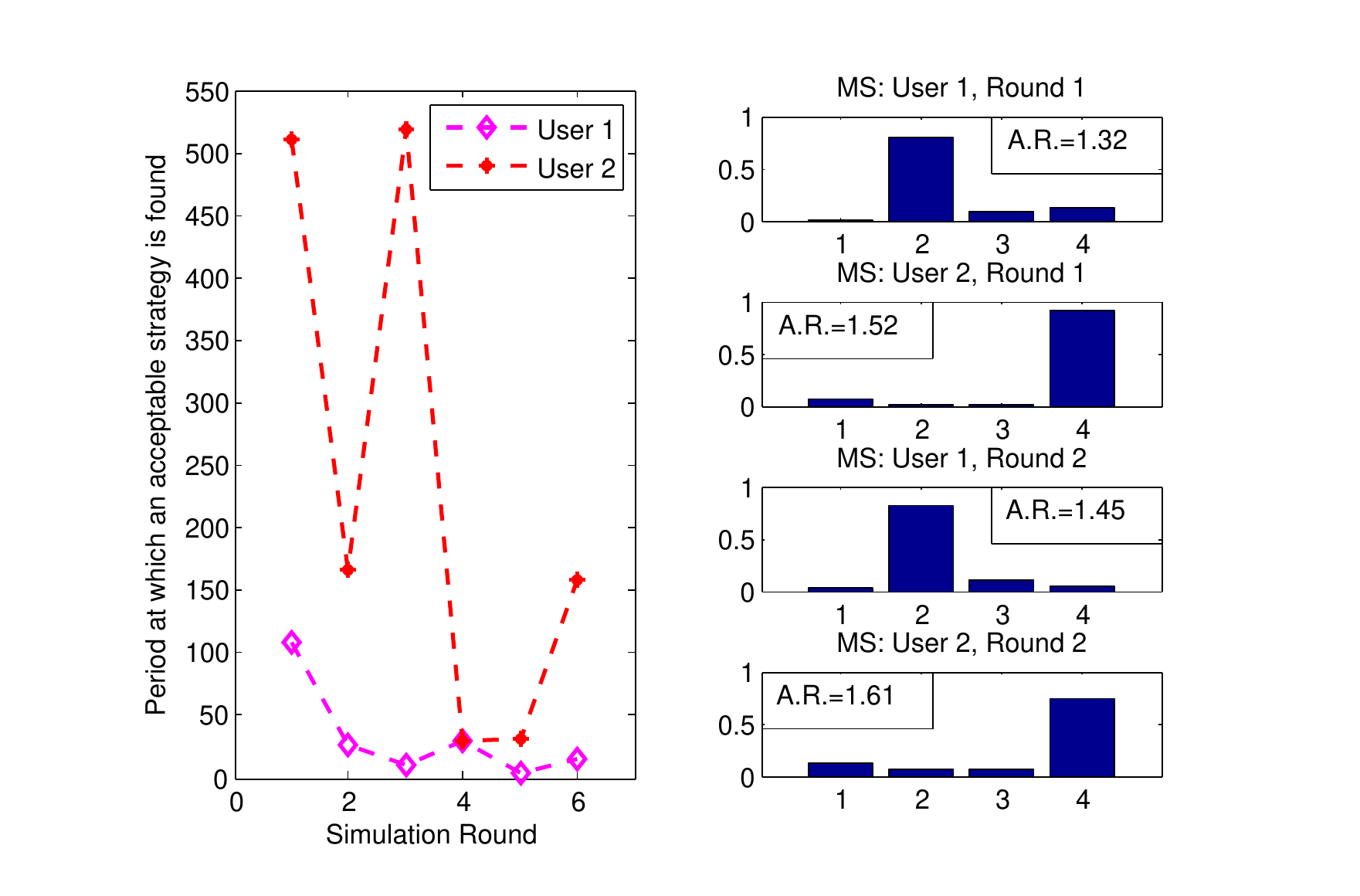}
\caption{Performance of BERTS. On the left, the vertical and horizontal axes show the 
periods and round number, respectively. The two curves depict the period at which a 
suitable mixed strategy (MS) is found at each of the 6 rounds. On the right, these mixed 
strategies are shown for both users at rounds 1 and 2, together with average rewards. 
The horizontal and vertical axes respectively depict the indices of actions and their 
selection probabilities.}
\label{BERT}
\end{figure}

The performance of BERTS, however, is not an explicit function of game duration. As described 
before, the procedure continues to search mixed strategies until a suitable one, which yields 
a regret less than the selected threshold, is captured. Then this strategy is played for the 
rest of the game. Theorem 7.8 of \cite{Bianchi06} specifies the minimum game duration to guarantee the 
convergence of BERTS, which is relatively long even for small number of users and actions. 
Nevertheless, similar to other search-based algorithms, there also exists the possibility of 
finding some acceptable strategy at early stages of the game. As a result, for relatively short 
games, the performance of BERTS is rather unpredictable. The other issue is the effect of regret 
threshold. On the one hand, larger threshold reduces the search time, since the set of acceptable 
strategies is large. On the other hand, large regret threshold might lead to performance 
loss, since there is the possibility that the user gets locked at some sub-optimal strategy 
at early stages, thereby incurring large accumulated regret. It is worth noting that due to its 
simplicity, and despite unpredictable performance, BERTS is an appealing approach in cases 
where computational effort should be minimized, and convergence to Nash equilibrium is desired.
Figure \ref{BERT} summarizes the results of few exemplary performances of BERTS. The parameters 
are selected as $T=80$, $M=1500$ and $\rho=0.16$ (see Section \ref{sec:BERT}). Simulation is 
performed for six \textit{independent} rounds. The curve on the left side of Figure \ref{BERT} 
depicts the period ($1\leq m\leq 1500$) at which the algorithm finds an acceptable strategy. As 
expected, the results exhibit no specific pattern. The four sub-figures on the right depict the mixed 
strategies selected by BERTS at rounds 1 and 2, together with average rewards. From this figure, 
at round 2, acceptable strategies are found earlier than round 1 by both users, leading to better 
average performance. It is also worth noting that for User 2, the strategy of round 1 is in essence 
better than that of round 2; nevertheless, it is found later. As a result, the average performance 
of round 2 is superior to that of round 1. 

\subsection{Part Two}\label{subsec:NumericTwo}
In this section we consider a wireless network consisting of 5 users (transmitter-receiver pairs), that 
compete for access to three orthogonal channels at two possible power levels (hence six actions). We 
compare BFPLS and BEWAS with the following selection approaches.\footnote{As mentioned before, observing 
the joint action profile and/or communication among users is not required for implementing BEWAS, BFPLS 
and BERTS. Therefore, they cannot be compared with strategies that include mutual observation and/or 
communication. A good example of such algorithms is the widely-used \textit{best-response dynamics}, 
where the strategy of each player is to play with the best-response to either the historical \cite{Chen11} 
or the predicted \cite{Maghsudi13} joint action profile of opponents. Another example is the strategy suggested 
in \cite{KalathilD12}, which is a combination of learning and auction algorithms where users communicate 
with each other.}
\begin{itemize}
\item Optimal (centralized) action assignment as described in Section \ref{subsec:Discussion}.
\item Centralized no-collision action selection, where no reward is assigned to users that 
access the same channel. Thus, users are encouraged to avoid collisions (a collision-avoidance 
strategy). This curve can be considered as an upper-bound for the performance of learning 
algorithms that select actions based on collision avoidance, such as \cite{KalathilD12}.
\item $\epsilon$-greedy algorithm, where at each trial, with probability $\epsilon$ (exploration 
parameter), an action is selected uniformly at random, while with probability $1-\epsilon$ the 
best action so far is played. The average reward of selected action is updated after each play 
\cite{Nie99}. For stationary environments, $\epsilon$ is usually time-varying and converges to 
zero in the limit, while in adversarial cases, $\epsilon$ is preferred to remain fixed. Here we 
let $\epsilon=0.1$.
\item Greedy approach, where at the beginning of the game, some trials are reserved for exploration, 
in which actions are selected at random (exploration period). The length of this period is a pre-defined 
fraction of the entire game duration. Based on the rewards of exploration period, the best possible 
action is selected, and is played for the rest of the game (exploitation period) \cite{Bianchi06}. 
This approach is extremely simple to implement; however, to the best of our knowledge, there is no 
analysis on the optimal length of the exploration period. 
\item Uniformly random selection.
\end{itemize} 
\begin{figure}[t]
\centering
\includegraphics[width=0.7\textwidth]{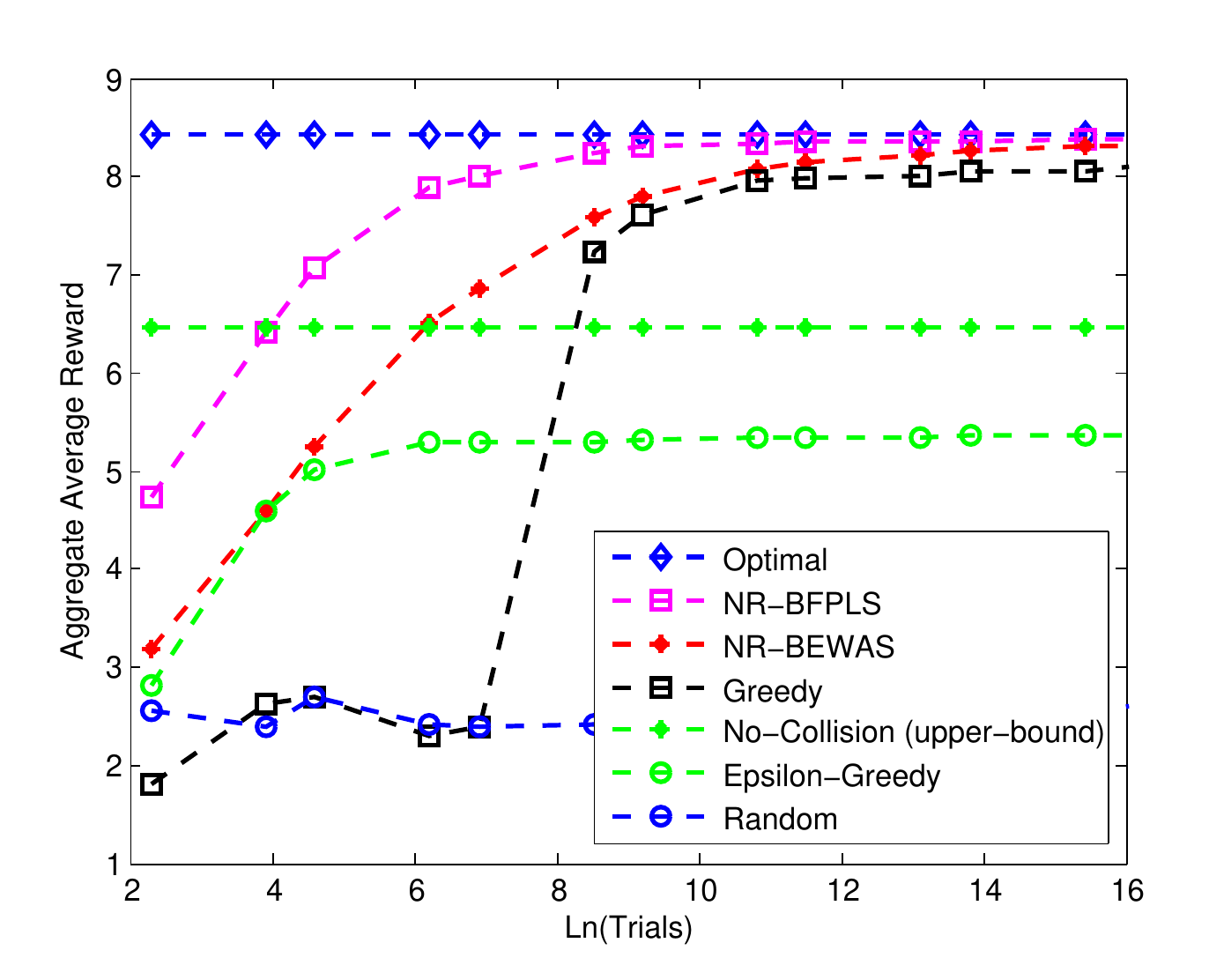}
\caption{Aggregate average reward of BFPLS and BEWAS compared to some other selection strategies.}
\label{fig:comp}
\end{figure}
The numerical results are depicted in Figure \ref{fig:comp}. From this figure, we can conclude the
following.
\begin{itemize}
\item The performance of interference-avoidance strategies is strongly influenced by channel matrices 
and tends to be poor specifically when the number of channels is less than that of users. The reason 
is that the sum reward of multiple interfering users with limited transmit power might be larger 
than the maximum achievable reward of any single user.
\item The performance of both BFPLS and BEWAS converge to that of centralized approach. As expected, 
BFPLS converges faster than BEWAS and we point out that the convergence speed of both algorithms would 
be dramatically enhanced if some side information was available to players, e.g. if users observed the 
actions of each other, or if communication was allowed among players. It is also worth noting that 
although BFPLS converges faster than BEWAS, the computation of integral (\ref{eq:ProbCal}) might be 
involved, especially for large number of actions \cite{Hutter05}.
\item In general, $\epsilon$-greedy and greedy approaches can be implemented easily with low computational 
cost; nevertheless, it can be seen that the greedy approaches are inferior to BEWAS and BFPLS in terms of 
asymptotic performance. Basically, these approaches are more suitable for stationary environments.
\end{itemize}
\section{Conclusion and Remarks}\label{sec:conclusion}
This paper deals with resource allocation in multi-user infrastructureless wireless networks. 
The problem of utility maximization has been formulated using the multi-player multi-armed 
bandit theory framework. More precisely, given no side information, the users aim at minimizing 
some regret expressed in terms of the loss of reward by selecting appropriate actions on a given 
space of transmit power levels and orthogonal frequency channels. Based on some recent mathematical 
results, we have designed two selection strategies, which not only provide vanishing regret for 
each player, but also guarantee the asymptotic convergence of the game to the set of correlated 
equilibria. We have also studied experimental regret testing strategy that asymptotically converges 
to the set of Nash equilibria. Numerical results confirms the applicability of the game model 
and proposed strategies to wireless channel selection and power control.

\section{Appendix}\label{sec:Appendix}
\subsection{Some Auxiliary Results}\label{subsec:auxiliary}
In this section, we state some auxiliary results and materials from game theory as well 
as bandit theory that are necessary for proofs. 
\subsubsection{Game Theory}\label{subsubsec:gameAux}
Throughout this part, we consider a game $\mathfrak{G}$ consisting of a set of $K$ players 
where the strategy set of each player $k \in \left \{ 1,...,K \right \}$ is denoted by 
$I^{(k)}$ with a generic element $i^{(k)}=(i_{1}^{(k)},...,i_{M}^{(k)})$. Similarly, the 
set of joint strategy profiles of players is denoted by $\mathbf{I}$ with a generic element 
$\mathbf{i}=(i^{(1)},...,i^{(K)})$ and $\mathbf{i}_{k}^{-}$ stands for the joint action 
profile of all players except for player $k$. Moreover, $g^{(k)}(\mathbf{i})$ stands for 
the utility function of some player $k$.\footnote{Note that compared to the system model 
some notation has been changed slightly.}
\begin{definition}
\label{de:smooth}
A game $\mathfrak{G}$ is smooth if, for each $k \in \left \{1,...,K \right \}$, $g^{(k)}(\mathbf{i})$ 
has continuous partial derivatives with respect to the components of $i^{(k)}$. 
\end{definition}
\begin{definition}
\label{de:strMonPayoff}
Let $\triangledown g^{(k)}=\left ( \frac{\partial g^{(k)}}{\partial i_{1}^{(k)}},\cdots, 
\frac{\partial g^{(k)}}{\partial i_{M}^{(k)}}\right )$, and call $\left (\triangledown g^{(k)}\right )_{k\in \left \{ 1,...,K \right \}}$ 
the payoff gradient of a smooth game $\mathfrak{G}$. We say that the payoff gradient is 
strictly monotone if 
\begin{equation}
\sum_{k=1}^{K} \left ( \triangledown g^{(k)}(\mathbf{i})- \triangledown g^{(k)}(\mathbf{j})\right )^{T}\left ( i^{(k)}-j^{(k)} \right )<0 ,
\end{equation} 
holds for all $\mathbf{i},\mathbf{j} \in \mathbf{I}$ with $\mathbf{i} \neq \mathbf{j}$.
\end{definition}
\begin{theorem}[\cite{Ui08}]
\label{th:UniqUi}
Consider a smooth game $\mathfrak{G}$ with compact strategy sets. If the payoff gradient of $\mathfrak{G}$ 
is strictly monotone then it has a unique correlated equilibrium, which places probability one on a unique 
pure-strategy Nash equilibrium.
\end{theorem}
\begin{definition}
\label{de:PotentialGame}
A game $\mathfrak{G}$ is potential if there exists a potential function $f: \mathbf{I} \to \Re$ such that 
\begin{equation}
\label{eq:potCond}
g^{(k)}(i,\mathbf{i}_{k}^{-})-g^{(k)}(j,\mathbf{i}_{k}^{-})= f(i,\mathbf{i}_{k}^{-})-f(j,\mathbf{i}_{k}^{-}),
\end{equation}
for all $i,j \in I^{(k)}$  and $k \in \left \{1,...,K \right \}$. 
\end{definition}
\begin{theorem}[\cite{Neyman97}]
\label{th:PotentialMax}
Let $\mathfrak{G}$ be a smooth potential game with a strictly concave potential function. Then a strategy 
profile is the unique pure strategy Nash equilibrium if and only if it is the potential maximizer.
\end{theorem}
\begin{lemma}[\cite{Ui08}]
\label{lm:PotenCon}
Let $\mathfrak{G}$ be a smooth potential game. A potential of $\mathfrak{G}$ is strictly concave if and only 
if the payoff gradient of $\mathfrak{G}$ is strictly monotone.   
\end{lemma}
\subsubsection{Bandit Theory}\label{subsubsec:banditAux}
\begin{lemma}
\label{lm:LemmaOne}
Let $R_{n}$ and $R_{\textup{Ext}}$ be given by (\ref{eq:CumRegret}) and (\ref{eq:external}), respectively. 
Then, for any $\delta \in (0,\frac{1}{2}]$, we have\footnote{Throughout this section and in order to simplify 
the notation, the player index ($k$) is omitted unless ambiguity arises.}
\begin{equation}
\textup{Pr}\left (\left |R_{n}-R_{\textup{Ext}} \right |\leq \sqrt{\frac{n}{2}\ln\frac{1}{\delta}} \right)\geq 1-2\delta,
\end{equation}
from which it follows that if $R_{n} \in o(n)$, then we have $R_{\textup{Ext}}\in o(n)$, with arbitrarily high 
probability.\footnote{Here and hereafter, the statement ''$X(n)\in o(n)$ with arbitrarily high probability'' 
for some nonnegative random sequence $X(n)\in\mathbb{R}$ means that the probability of $X(n)\notin o(n)$ can 
be made arbitrarily small, provided that some parameter is chosen sufficiently small.}
\end{lemma}
\begin{IEEEproof}
By comparing (\ref{eq:CumRegret}) and (\ref{eq:external}), it suffices to show that 
$\textup{Pr}\left (\left |\sum_{t=1}^{n}\left (g_{t}(I_{t})-\bar{g}_{t}
(\mathbf{P}_{t}) \right)\right| \right ) \leq \sqrt{\frac{n}{2}\ln\frac{1}{\delta}}\leq 1-2\delta $. 
To this end, define $S :=\sum_{t=1}^{n} g_{t}(I_{t})$, where $g_{t}(I_{t}) \in [0,1]$, 
$1\leq t \leq n$, are independent random variables (see also Section \ref{subsec:Regret}). 
Further note that $\bar{S}=\textup{E}[S]=\sum_{t=1}^{n}\bar{g}_{t}(\mathbf{P}_{t})$. 
Therefore, by Hoeffding's inequality \cite{Bianchi06},
\begin{equation}
\begin{aligned}
\textup{Pr}\left(\left|R_{n}-R_{\textup{Ext}} \right |\geq \sqrt{\frac{n}{2}\ln\frac{1}{\delta}} \right)=
&\textup{Pr}\left (\left |S-\bar{S} \right | \geq  \sqrt{\frac{n}{2}\ln\frac{1}{\delta}}\right) \\
\leq & 2 \exp \left (-\frac{2\frac{n}{2}\ln\frac{1}{\delta}}{n}\right)=2\delta.
\end{aligned}
\end{equation}
Hence the Lemma follows with
$\textup{Pr}\left(\left|R_{n}-R_{\textup{Ext}} \right| \leq \sqrt{\frac{n}{2}\ln\frac{1}{\delta}} \right)=
1-\textup{Pr}\left(\left|R_{n}-R_{\textup{Ext}} \right |\geq \sqrt{\frac{n}{2}\ln\frac{1}{\delta}} \right)$.
\end{IEEEproof}
\begin{lemma}
\label{lm:LemmaTwo}
Let $R_{\textup{Ext}}$ be given by (\ref{eq:external}). Moreover, define $\tilde{R}_{n}=\max_{i=1,...,N}\sum_{t=1}^{n}g_{t}(i)-\sum_{t=1}^{n}\tilde{g}_{t}\left(\mathbf{P}_{t}\right)$,
where $\tilde{g}_{t}\left (\mathbf{P}_{t} \right)=\sum_{i=1}^{N}p_{i,t}\tilde{g}_{t}(i)$ and $\tilde{g}_{t}(i)$ 
is given by (\ref{eq:GainEstim}). Then we have
\begin{equation}
\textup{Pr}\left (\left |\tilde{R}_{n}-R_{\textup{Ext}}\right |\leq \sqrt{\frac{n}{2}\ln\frac{1}{\delta}}\right)\geq 1-2\delta.
\end{equation}
Hence, for sufficiently small $\delta>0$, $R_{\textup{Ext}} \in o(n)$
implies that $\tilde{R}_{n} \in o(n)$, with arbitrarily high probability.
\end{lemma}
\begin{IEEEproof}
Similar to the proof of Lemma \ref{lm:LemmaOne}, it follows from (\ref{eq:external}) 
and the definition of $\tilde{R}_{n}$ that it is sufficient to show that $\textup{Pr}
\left (\left |\sum_{t=1}^{n}\left (\tilde{g}_{t}(\mathbf{P}_{t})-\bar{g}_{t}
(\mathbf{P}_{t})\right)\right | \leq \sqrt{\frac{n}{2}\ln\frac{1}{\delta}} 
\right)\leq 1-2\delta$ for $\delta \in (0,\frac{1}{2}]$. To this end, note that 
$\tilde{g}_{t}(\mathbf{P}_{t}) \in [0,1]$, $1\leq t \leq n$, are independent 
random variables. Moreover, since $\tilde{g}_{t}(i)$ 
is an unbiased estimate of $g_{t}(i)$, we have $\textup{E}[\sum_{t=1}^{n} \tilde{g}_{t}(\mathbf{P}_{t})]
=\sum_{t=1}^{n}\bar{g}_{t}(\mathbf{P}_{t})$. Hence, defining $S=\bar{g}_{t}(\mathbf{P}_{t})
-\tilde{g}_{t}(\mathbf{P}_{t})$ and proceeding as in the proof of Lemma \ref{lm:LemmaOne} 
with the Hoeffding's inequality in hand proves the lemma.
\end{IEEEproof}
\begin{proposition}
\label{th:Regret_EstRegret}
Let $R_{n}$ be given by (\ref{eq:CumRegret}) and $\tilde{R}_{n}$ be defined as in Lemma \ref{lm:LemmaTwo}. 
Then, $R_{n}\in o(n) $ implies that$\tilde{R}_{n} \in o(n)$.
\end{proposition}

\begin{IEEEproof}
  Lemma \ref{lm:LemmaOne} implies that $R_{n} \in o(n)\Rightarrow
  R_{\textup{Ext}} \in o(n)$ with arbitrarily high probability, while
  by Lemma \ref{lm:LemmaTwo}, we
  have $R_{\textup{Ext}} \in o(n) \Rightarrow \tilde{R} \in
  o(n)$. Therefore, if $R_{n} \in o(n)$, then $\tilde{R}\in o(n)$ with
  arbitrarily high probability.
\end{IEEEproof}
\begin{theorem}(\cite{Bianchi06})
\label{th:RegretBound}
Let $\Phi (\mathbf{U})=\psi (\sum_{i=1}^{N}\phi (u_{i}))$, 
where $\textbf{U}=(u_{1},...,u_{N})$. Consider a selection strategy, which at time 
$t$ selects action $I_{t}$ according to distribution $\mathbf{P}_{t}$, whose elements 
$p_{i,t}$ are defined as
\begin{equation}
p_{i,t}=(1-\gamma_{t})\frac{\phi'(R_{i,t-1})}{\sum_{k=1}^{N}\phi'(R_{i,t-1})}+\frac{\gamma_{t}}{N},
\end{equation}
where $R_{i,t-1}=\sum_{s=1}^{t-1}\left(g_{s}(i)-g_{s}(I_{s})\right)$. 
Assume that:\\
A1. $\sum_{t=1}^{n}\frac{1}{\gamma_{t}^{2}}=o(\frac{n^{2}}{\ln n})$,\\
A2. For all vectors $\mathbf{V}_{t}=(v_{1,t},...,v_{n,t})$ with $\left |v_{i,t} \right|\leq \frac{N}{\gamma_{t}}$, 
we have 
\begin{equation}
\label{eq:ATwo}
\lim_{n \to \infty }\frac{1}{\psi (\phi (n))}\sum_{t=1}^{n}C(\mathbf{V}_{t})=0,
\end{equation}
where $C(\mathbf{V}_{t})= \sup_{\mathbf{U}\in \mathbb{R}^{N}}{\psi}'(\sum_{i=1}^{N}\phi (u_{i}))\sum_{i=1}^{N}{\phi}'' (u_{i})v_{i,t}^{2}$.\\
A3. For all vectors $\mathbf{U_{t}}=(u_{1,t},...,u_{n,t})$, with $u_{i,t}\leq t$, 
\begin{equation}
\label{eq:AThree}
\lim_{n \to \infty }\frac{1}{\psi (\phi (n))}\sum_{t=1}^{n}\gamma _{t}\sum_{i=1}^{N}\triangledown_{i}\Phi (\mathbf{U}_{t})=0.
\end{equation}
A4. For all vectors $\mathbf{U_{t}}=(u_{1,t},...,u_{n,t})$, with $u_{i,t}\leq t$,
\begin{equation}
\label{eq:AFour}
\lim_{n \to \infty }\frac{\ln n}{\psi (\phi (n))}\sqrt{\sum_{t=1}^{n}\frac{1}{\gamma_{t}^{2}}\left(\sum_{i=1}^{N}\triangledown_{i}\Phi (\mathbf{U}_{t})\right)^{2}}.
\end{equation}
Then the selection strategy satisfies
\begin{equation}
\label{eq:Result}
\lim_{n \to \infty}\frac{1}{n}\left (\max_{i=1,...,N} \sum_{t=1}^{n}g_{t}(i)- \sum_{t=1}^{n}g_{t}(I_{t}) \right)=0,
\end{equation}
or equivalently, $R_{n} \in o(n)$, where $R_{n}$ is given by (\ref{eq:CumRegret}).
\end{theorem}
%


%
\subsection{Proof of Proposition \ref{pr:EqType}}\label{subsec:CorEq-Pr}
In order to prove Proposition \ref{pr:EqType}, we use Theorem
\ref{th:UniqUi}. As the strategy set is compact, in order to use this
theorem, we show that 1) the game is smooth, and 2) the payoff
gradient is strictly monotone.

According to our system model, by changing the channel index,
$c^{(k)}$, the channel gain and interference changes. Therefore we
define $i_{1}^{(k)}:=\frac{|h_{kk',c^{(k)}}|^{2}} {\sum_{q=1}^{ Q_{k}}
  l^{(q)}|h_{qk',c^{(k)}}|^{2} + N_{0}} $ and
$i_{2}^{(k)}:=l^{(k)}$, from which we have $g^{(k)}(\mathbf{i}) =
\log\bigl(i_{1}^{(k)}i_{2}^{(k)}\bigr)-\alpha i_{2}^{(k)}$. This
results in
\begin{equation}
\frac{\partial g^{(k)}}{\partial i_{1}^{(k)}}=\frac{1}{i_{1}^{(k)}}.
\end{equation}
and 
\begin{equation}
\frac{\partial g^{(k)}}{\partial i_{2}^{(k)}}=\frac{1}{i_{2}^{(k)}}-\alpha.
\end{equation}
Hence by Definition \ref{de:smooth}, the game is smooth. On the other hand,
given $g^{(k)}$, we have
\begin{equation}
\begin{aligned}
 \left ( \triangledown g^{(k)}(\textbf{i})- \triangledown g^{(k)}(\textbf{j})\right )^{T}
 \left ( i^{(k)}-j^{(k)} \right )~~~~~~~~~~~~~~~& \\ 
 =\begin{bmatrix}
\frac{1}{i_{1}^{(k)}}-\frac{1}{j_{1}^{(k)}} & \frac{1}{i_{2}^{(k)}}-\frac{1}{j_{2}^{(k)}}
\end{bmatrix}\begin{bmatrix}
i_{1}^{(k)}-j_{1}^{(k)}\\i_{2}^{(k)}-j_{2}^{(k)} 
\end{bmatrix}~~~~~~~~~& \\ 
= \left (\frac{1}{i_{1}^{(k)}}-\frac{1}{j_{1}^{(k)}}  \right )(i_{1}^{(k)}-j_{1}^{(k)})+
\left (\frac{1}{i_{2}^{(k)}}-\frac{1}{j_{2}^{(k)}} \right )&(i_{2}^{(k)}-j_{2}^{(k)}),
\end{aligned}
\end{equation}
which is always negative as for any $x,y>0$ and $x \neq y$, $x-y>0$
yields $\frac{1}{x}-\frac{1}{y}<0$ and vice versa. So,
\begin{equation}
\sum_{k=1}^{K}\triangledown g^{(k)}< 0
\end{equation}
i.e. the payoff gradient is strictly monotone by Definition
\ref{de:strMonPayoff}.

As a result, by Theorem \ref{th:UniqUi}, the game has a unique
correlated equilibrium which places probability one on the unique Nash
equilibrium.

\subsection{Proof of Proposition \ref{pr:EqTypeTwo}}\label{subsec:CorEq-Pr-Two}
%
%
%

First, we point out that the game is a potential game with a potential
function being $f(\cdot)=\sum_{k=1}^{K}g^{(k)}(\cdot)$, by simply
inserting $g^{(k)}$ and $f$ in condition (\ref{eq:potCond}). Moreover, 
similar to Proposition \ref{pr:EqType}, it can be easily shown that the 
game is smooth and the payoff gradient is strictly monotone (Define 
$i_{1}^{(k)}:=\frac{\left | h_{c^{(k)}}\right |^{2}}{N_{0}}$ and 
$i_{2}^{(k)}=l^{(k)}$).\footnote{Details are similar to Proposition 
\ref{pr:EqType} and thus are omitted.}~Therefore, by Lemma \ref{lm:PotenCon}, 
the game is a smooth potential game with strictly concave potential 
function. As a result, by Theorem \ref{th:PotentialMax}, it has a unique 
pure strategy Nash equilibrium which is the potential maximizer. On 
the other hand, by Lemma \ref{th:UniqUi}, the game has a unique 
correlated equilibrium which places probability one on the unique 
pure strategy Nash equilibrium.

\subsection{Proof of Proposition \ref{pr:NR-BEWAS-One}}\label{subsec:NR-BEWAS-Pr-One}
We first notice that $R_{\textup{Ext}}\in O((nN)^{\frac{2}{3}}(\ln N)^{\frac{1}{3}})$, as 
stated by the following lemma.\footnote{Throughout this section and in order to simplify 
the notation, the player index ($k$) is omitted unless ambiguity arises.}
\begin{lemma}
\label{lm:pr-BEWAS-One}
Consider a selection strategy that uses $\mathbf{P}_{t}=\left(p_{1,t},...,p_{N,t}\right)$ 
to select an action among $N$ possible choices, where $p_{i,t}$ is calculated as
\begin{equation}
\label{eq:pr-BEWASO}
p_{i,t}=(1-\gamma)\frac{\exp(\eta\tilde{R}_{i,t-1})}{\sum_{m=1,..,N}\exp(\eta\tilde{R}_{m,t-1})}+\frac{\gamma}{N},
\end{equation}
and $\tilde{R}_{i,t-1}$ denotes the estimated accumulated regret of not playing action $i$.\footnote{This 
definition should not be mistaken for the general regret defined in Section \ref{subsec:Regret}.}~Then 
selecting $\gamma$ and $\eta$ as given by Proposition \ref{pr:NR-BEWAS-One} yields $R_{\textup{Ext}}
\in O((nN)^{\frac{2}{3}}(\ln N)^{\frac{1}{3}})$. 
\end{lemma}
\begin{IEEEproof}
The proof is a direct corollary of Theorem 6.6 of \cite{Bianchi06}.
\end{IEEEproof}
Given Lemma \ref{lm:pr-BEWAS-One}, we follow the approach of \cite{Stoltz05} for the rest of the proof.

Recall that by Section \ref{subsec:Conversion} and Algorithm \ref{alg:NR-BEWAS}, the mixed strategy of each 
player is defined by
\begin{equation} 
\label{eq:deltaProbC} 
\mathbf{P}_{t}=\sum_{(i \to j):i\neq j}\mathbf{P}_{t}^{(i \to j)}\delta_{(i \to j),t}.
\end{equation}
Hence,
\begin{equation} 
\label{eq:theorem}
\bar{g}_{t}(\mathbf{P}_{t})=\sum_{(i \to j):i\neq j}\bar{g}_{t}(\mathbf{P}_{t}^{(i\to j)})\delta_{(i \to j),t}.
\end{equation}
Lemma \ref{lm:pr-BEWAS-One} specifies the growth rate of external regret. On the other hand, as described in 
Section \ref{subsec:Conversion}, the convergence approach applies the BEWAS algorithm for $N(N-1)\leq N^2$ 
actions. Therefore, (\ref{eq:theorem}) together with Lemma \ref{lm:pr-BEWAS-One} yields 
\begin{equation} 
\max \sum_{t=1}^{n}\bar{g}_{t}(\mathbf{P}_{t}^{(i\to j)})-\sum_{t=1}^{n} \bar{g}_{t}(\mathbf{P}_{t}) \in O((N^{2}n)^{\frac{2}{3}}
(2\ln N)^{\frac{1}{3}}),
\end{equation}
and the definition of internal regret ensures that $\max_{i\neq j}R_{(i \to j),n} \in O((N^{2}n)^{\frac{2}{3}}
(\ln N)^{\frac{1}{3}})$, which concludes the proof. Details can be found in \cite{Stoltz05}, and hence are omitted.
\subsection{Proof of Proposition \ref{pr:NR-BEWAS}}\label{subsec:NR-BEWAS-Pr}
We first show that the algorithm has vanishing external regret, i.e. $R_{\textup{Ext}}\in o(n)$, 
as formalized in the following.
\begin{lemma}
\label{lm:pr-BEWAS}
Consider a selection strategy that uses $\mathbf{P}_{t}=\left(p_{1,t},...,p_{N,t}\right)$ 
to select an action among $N$ possible choices, where $p_{i,t}$ is calculated as
\begin{equation}
\label{eq:pr-BEWAS}
p_{i,t}=(1-\gamma_{t})\frac{\exp(\eta_{t}\tilde{R}_{i,t-1})}{\sum_{m=1,..,N}\exp(\eta_{t}\tilde{R}_{m,t-1})}+\frac{\gamma_{t}}{N},
\end{equation}
and $\tilde{R}_{i,t-1}$ denotes the estimated accumulated regret of not playing action $i$. Then, 
for $\gamma_{t}$ and $\eta_{t}$ as given by Proposition \ref{pr:NR-BEWAS}, this strategy yields vanishing 
external regret, i.e. $R_{\textup{Ext}}\in o(n)$. 
\end{lemma}
\begin{IEEEproof}
By Proposition \ref{th:Regret_EstRegret}, if (\ref{eq:Result})
is satisfied for a selection strategy (that is, if $R_{n} \in o(n)$),
then the growth rate of the external regret caused by the bandit
version of that strategy (which uses estimated rewards instead of true
ones) grows sublinearly in $n$, i.e. $\tilde{R}_{n} \in o(n)$. Therefore,
in order to prove the proposition, we can show that our selected
parameters $\gamma_{t}=t^{-\frac{1}{3}}$ and $\eta
_{t}=\frac{\gamma_{t}^{3}}{N^{2}}$ satisfy axioms A1-A4 of Theorem
\ref{th:RegretBound}. In what follows, we show that each of these
axioms is fulfilled. In doing so, we omit for the lack of space simple
calculus steps. Also the reader should note that in our strategy we
have $\Phi (\textbf{U})=\frac{1}{\eta_{t}}\ln\left (\sum_{i=1}^{N}
  \exp(\eta _{t}u_{i})\right)$.
\begin{itemize}
\item[A1.] For $\gamma_{t}=t^{-\frac{1}{3}}$, we have
\begin{equation}
\label{eq:AOnePrim}
\sum_{t=1}^{n}\frac{1}{\gamma_{t}^{2}}=\sum_{t=1}^{n}t^{\frac{2}{3}}=\textup{Harmonic Number}[n,-\frac{2}{3}]:=\textup{H}_{n}[\frac{-2}{3}].
\end{equation}
Then, 
\begin{equation}
\lim_{n \to \infty} \frac{\ln n}{n^{2}}\sum_{t=1}^{n}\gamma _{t}^{2}= \lim_{n \to \infty}\frac{\ln n}{n^{2}}\textup{H}_{n}[\frac{-2}{3}]=0.
\end{equation}
%
%
\item[A2.] For $\psi (x)=\frac{1}{\eta _{t}}\ln x$ and $\phi (x)=\exp(\eta _{t}x)$, we obtain
\begin{equation}
\label{eq:ATwoPrim}
C(\mathbf{V}_{t})=\sup \left( \eta _{t}\sum_{i=1}^{N} v_{i,t}^{2} \right)=\frac{\eta _{t}N^{3}}{\gamma_{t}^{2}}.
\end{equation}
Hence,
\begin{equation}
\begin{aligned}
\lim_{n \to \infty }\frac{1}{\psi (\phi (n))}\sum_{t=1}^{n}&C(\mathbf{V}_{t})= \lim_{n \to \infty }\frac{1}{n}\sum_{t=1}^{n} t^{\frac{-1}{3}}  \\ 
=\lim_{n \to \infty }\frac{1}{n}\textup{H}_{n} [\frac{1}{3}]=&0.
\end{aligned}
\end{equation}
%
%
\item[A3.] For $\Phi (\textbf{U})=\frac{1}{\eta _{t}}\ln\left (\sum_{i=1}^{N} \exp(\eta _{t}u_{i})\right )$, 
$\triangledown_{i}\Phi (\mathbf{U}_{t})$ yields
\begin{equation}
\label{eq:AThreePrim}
\triangledown_{i}\Phi (\mathbf{U}_{t})=\frac{\exp(\eta_{t}u_{i})}{\sum_{i=1}^{N}\exp(\eta_{t}u_{i})}.
\end{equation}
Therefore,
\begin{equation}
\begin{aligned}
  \lim_{n \to \infty }\frac{1}{\psi (\phi (n))}~&\sum_{t=1}^{n}\gamma _{t}\sum_{i=1}^{N}\triangledown_{i}\Phi (\mathbf{U}_{t})=\\
  \lim_{n \to \infty }\frac{1}{n}\sum_{t=1}^{n}t^{\frac{-1}{3}}&\sum_{i=1}^{N} \frac{\exp(\eta_{t}u_{i})}{\sum_{i=1}^{N}\exp(\eta_{t}u_{i})}=\\
  \lim_{n \to \infty }\frac{1}{n}~\textup{H}_{n}[\frac{1}{3}]&=0.
\end{aligned}
\end{equation}
\item[A4.] $A4$ follows simply by substituting (\ref{eq:AThreePrim}) in (\ref{eq:AFour}). 
\end{itemize}

Hence, all axioms A1-A4 are satisfied, and therefore (\ref{eq:Result}) holds, which, 
together with Proposition \ref{th:Regret_EstRegret}, completes the proof.
\end{IEEEproof}

By Lemma \ref{lm:pr-BEWAS}, the external regret BEWAS grows sublinearly in $n$. Therefore, similar to 
the proof of Proposition \ref{pr:NR-BEWAS-One}, (\ref{eq:theorem}) yields 
\begin{equation} 
\max \sum_{t=1}^{n}\bar{g}_{t}(\mathbf{P}_{t}^{(i\to j)})-\sum_{t=1}^{n} \bar{g}_{t}(\mathbf{P}_{t}) \in o(n),
\end{equation}
and the definition of internal regret ensures that $\max_{i\neq j}R_{(i \to j),n} \in o(n)$, 
which concludes the proof.

%

\bibliographystyle{IEEEbib}
\bibliography{Myreferences}
\end{document}